\documentclass[fleqn,usenatbib]{mnras}
\usepackage{newtxtext,newtxmath}

\usepackage[T1]{fontenc}
\usepackage{ae,aecompl}

\usepackage{mathtools}
\usepackage{adjustbox}
\usepackage{longtable}
\usepackage{float}

\usepackage{graphicx}	
\usepackage{amsmath}	
\usepackage{amssymb}	

\usepackage{pdflscape}
\usepackage{multirow}
\usepackage{multicol}
\usepackage{booktabs}

\makeatletter
\newcommand*\bigcdot{\mathpalette\bigcdot@{.5}}
\newcommand*\bigcdot@[2]{\mathbin{\vcenter{\hbox{\scalebox{#2}{$\m@th#1\bullet$}}}}}
\makeatother
\DeclareMathOperator*{\argmin}{arg\,min} 
\newcommand\teff{T$_{\mathrm{eff}}$}
\newcommand\logg{log\,{\it g}}
\newcommand\kms{km\,s$^{-1}$}

\newcommand{\qvec}[1]{\textbf{\textit{#1}}}

\title[Machine Learning on Stellar Spectra]{An Application of Deep Learning in the Analysis of Stellar Spectra}

\author[S. Fabbro et al.]
{S. Fabbro$^{1}$\thanks{E-mail: sebastien.fabbro@nrc-cnrc.gc.ca, kvenn@uvic.ca},
K.A. Venn$^{2}$,
T. O'Briain$^{1,2}$,
S. Bialek$^{1,2}$,
C.L. Kielty$^{2}$, 
F. Jahandar$^{2}$,
\newauthor
and S. Monty$^{2}$
\\
$^{1}$National Research Council of Canada, Herzberg Astronomy \& Astrophysics Program, 5071 West Saanich Road, Victoria BC, V9E 2E7, Canada \\
$^{2}$Department of Physics and Astronomy, University of Victoria, Victoria, BC, V8W 3P2 \\ }

\date{Accepted 19 December 2017. Received 6 August 2017.}

\pubyear{2017}

\begin{document}
\label{firstpage}
\pagerange{\pageref{firstpage}--\pageref{lastpage}}
\maketitle

\begin{abstract}

Spectroscopic surveys require fast and efficient analysis methods to maximize their scientific impact. Here we apply a deep neural network architecture to analyze both SDSS-III APOGEE DR13 and synthetic stellar spectra. 
When our convolutional neural network model (StarNet) is trained on APOGEE spectra, we show that the stellar parameters (temperature, gravity, and metallicity) are determined with similar precision and accuracy as the APOGEE pipeline.  StarNet can also predict stellar parameters when trained on synthetic data, with excellent precision and accuracy for both APOGEE data and synthetic data, over a wide range of signal-to-noise ratios. In addition, the statistical uncertainties in the stellar parameter determinations are comparable to the differences between the APOGEE pipeline results and those determined independently from optical spectra. We compare StarNet to other data-driven methods; for example, StarNet and the Cannon\,2 show similar behaviour when trained with the same datasets, however StarNet performs poorly on small training sets like those used by the original Cannon. The influence of the spectral features on the stellar parameters is examined via partial derivatives of the StarNet model results with respect to the input spectra.  While StarNet was developed using the APOGEE observed spectra and corresponding ASSET synthetic data, we suggest that this technique is applicable to other wavelength ranges and other spectral surveys.

\end{abstract}

\begin{keywords}
techniques: spectroscopic - methods: numerical - surveys - infrared: stars - stars: fundamental parameters
\end{keywords}



 \section{Introduction}
Spectroscopic surveys provide a homogeneous database of stellar spectra that are ideal for machine learning applications. A variety of techniques have been studied in the past two decades.  These range from the SDSS SEGUE stellar parameter pipeline using a decision tree architecture between spectral matching to a synthetic grid and line index measurements \citep[e.g.,][]{lee2008, allendeprieto2008, yanny2009segue, lee2011}, to the detailed algorithms employed for analysis of the Gaia spectroscopic data \citep[e.g.,][]{smiljanic2014gaiaeso, recioblanco2016, recioblanco2017, pancino2017}.

The use of artificial neural networks in astrophysical applications has a history going back more than 20 years, with pioneering research in stellar classification by authors such as \citet{von1994automated} and \citet{singh1998stellar}. In  \citet{bailer1997} and \citet{bailer2000stellar}, a neural network was applied to synthetic stellar spectra to predict the effective temperature \teff, surface gravity \logg, and metallicity [Fe/H]. 

More recently, dramatic improvements have occurred in the usability and performance of algorithms implemented in machine learning software. This, combined with the increase in computing power and the availability of large data sets, has led to the successful implementation of more complex neural network architectures. These have proven to be pivotal in difficult image recognition tasks and natural language processing. The earlier attempts at neural networks for stellar spectra analyses \citep[e.g.,][]{bailer2000stellar, manteiga2010anns} would train the neural network on synthetic spectra, and test the model on synthetic spectra. Machine learning methods were also used in one of the SEGUE pipelines \citep{lee2008}, where two neural networks were trained: one on synthetic spectra and the other on previous SEGUE parameters.

In this paper, we examine several cases of training on synthetic spectra and predicting stellar parameters on both synthetic and observed spectra, or training and predicting on only observed spectra in a purely data-driven approach.
The methods we use follow the supervised learning approach, where a representative subset of stellar spectra (either synthetic or observed) with known stellar parameters is selected. This subset can be further divided into a reference set, where the neural network learns the mapping function from spectra to stellar parameters, and a test set, used for ensuring the accuracy of the predictions. Once the network model is trained (i.e. with fixed network parameters), the stellar parameters can be predicted for the rest of the sample. The stellar parameters we consider for this spectral analysis are the effective temperature (\teff), surface gravity (\logg), and metallicity ([Fe/H]).

In this paper we present StarNet: a convolutional neural network model applied to the analysis of stellar spectra. We introduce our machine learning methods in Section 2, and evaluate our model for a set of synthetic data in Section 3.  As an exercise of its effectiveness, we apply StarNet to the APOGEE survey in Section 4 (DR13 and earlier data releases when appropriate), and compare the stellar parameters to those from the APOGEE pipeline(s). In Section 5, we discuss the success of our StarNet results to other stellar analyses, and confirm that neural networks can significantly increase the robustness, efficiency, and scientific impact of spectroscopic surveys.

\section{Machine Learning Methodology}

Supervised learning has been shown to be well adapted for continuous variable regression problems. Given a training set in which, for each input spectrum, there are known stellar parameters, a supervised learning model is then capable of approximating a function that transforms the input spectra to the output values. The learned function can then ideally be used to predict the output values of a different dataset. The particular form of this function and how it is learned depends on the neural network architecture. Summarized below is the convolutional neural network that we have implemented for the analysis of stellar spectra; we provide more details about deep neural networks and the mathematical operations of our selected architecture in Appendix \ref{appendix_method}.

\subsection {The StarNet convolutional neural network}
A neural network (NN) can be arranged in layers of artificial neurons: an \textit{input layer} (i.e. the input data), a number of \textit{hidden layers}, and an \textit{output layer}. Depending on the architecture of the NN, there may be several hidden layers, each composed of multiple neurons that weight and offset the outputs from the previous layer to compute input values for the following layer. The more hidden layers used in the NN, the \textit{deeper} the model is. The combination of these layers act as a single function, and in the case of StarNet, this function predicts three stellar parameters.

\begin{figure*}
\centering
\includegraphics[width=\linewidth]{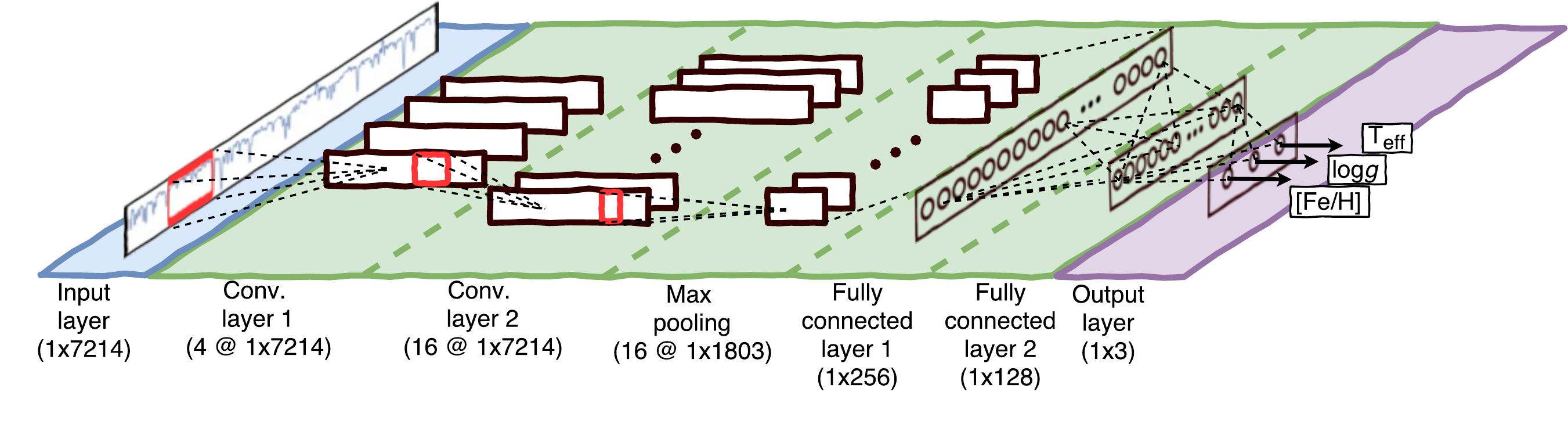}
\caption{
The current StarNet CNN model composed of seven layers. The first layer is solely the input data; followed by two convolutional layers with four and 16 filters (in successive order), then a max pooling layer with a window length of four units, followed by three fully connected layers with 256, 128, and three nodes (again, in successive order). The final layer is the output layer.
\label{fig:StarNet}}
\end{figure*}

Inspired from recent studies of deep neural networks on stellar spectra \citep{li2016dnn, wang2017spectra}, we have focused our analysis on deep architectures. The convolutional neural network (CNN) selected for StarNet is shown schematically in Fig.~\ref{fig:StarNet}. This architecture is composed of a combination of fully connected layers and 1-dimensional convolutional layers. Fully connected layers are the classical neural network layers that compute multiple linear combinations of all of the input values to produce an output vector. In the case of a convolutional layer, a series of filters are applied, extracting local information from the previous layer. Throughout the training phase the network learns the filters that are activated most strongly when detecting specific features, thus producing a collection of \textit{feature maps}. Using two successive convolutional layers,  the second of the two convolves across the previous layer's feature map, which allows the model to learn higher order features.

The combination of convolutional layers and fully connected layers in our StarNet implementation means that the output parameters are not only affected by individual features in the input spectrum, but also, combinations of features found in different areas of the spectrum are utilized. This technique strengthens the ability of StarNet to generalize its predictions on spectra with a wide range of signal-to-noise (S/N) ratios across a larger stellar parameter space. More details of the StarNet model itself are discussed in Section \ref{model_comparison}.

\subsection{Training and testing of the Model}
\label{training_method}
Before training the model, the reference set is split into a training set and a cross-validation set. Training is performed through a series of forward and back propagation iterations, using batches of the training set. The forward propagation is the model function itself: at each layer, weights are applied to all of the input values, and at the output layer, a prediction is computed. These predictions are then compared to the target values through a \textit{loss function}. In our StarNet model, a mean-squared-error loss function computes the loss between predictions and target values and a total loss for all training examples is minimized.

Initially, the weights of the model are randomly set and therefore the predictions will be quite poor. To improve these predictions, the model weights are updated following each batch forward propagation. Therefore, the weights are adjusted multiple times per \textit{iteration}.
A minimum was usually reached after 20-25 iterations (although this could vary greatly depending on the complexity of the model used). The training of our StarNet model reached convergence in an amount of time that depended on the size of the training set and the model architecture; for $\sim$41,000 spectra (close to 300 million intensity values) and ~7.5 million NN parameters, the training converged in 30 minutes, using a 16 cores virtual machine. 

When the training stage reaches convergence, the weight values are frozen. The estimated model is evaluated on a test set of spectra with a wide range of S/N. All tests of StarNet are quantified with location (mean, median) and spread (standard deviation, mean absolute difference) summary statistics on the residuals with a test set of known parameters.

\section {Synthetic Spectra}
\label{section:syntheticspectra}

\begin{table}
\centering
\caption{Stellar parameter distribution of the ASSET synthetic spectra grid}
\label{tab:ASSETgrid}
\resizebox{\columnwidth}{!}{
\begin{tabular}{clcccccc}
\hline
\multicolumn{1}{l}{\textbf{Class}} &  & \multicolumn{1}{l}\textbf{\teff} & \textbf{\logg} & \textbf{{[}M/H{]}} & \textbf{{[}C/M{]}} & \textbf{{[}N/M{]}} & \textbf{{[}$\alpha$/M{]}} \\ \hline
\multirow{3}{*}{GK} & Min. & 3500 & 0 & -2.5 & -1 & -1 & -1 \\ [.2ex]
 & Max. & 6000 & 5 & 0.5 & 1 & 1 & 1 \\
 & Step & 250 & 0.5 & 0.5 & 0.25 & 0.5 & 0.25 \\ \hline
\multirow{3}{*}{F} & Min. & 5500 & 1 & -2.5 & -1 & -1 & -1 \\ [.2ex]
 & Max. & 8000 & 5 & 0.5 & 1 & 1 & 1 \\
 & Step & 250 & 0.5 & 0.5 & 0.25 & 0.55 & 0.25 \\ \hline
\end{tabular}}
\end{table}

As mentioned in Section 1, machine learning techniques have been applied to stellar spectra in the past using synthetic data for training and testing \citep{bailer2000stellar}. Synthetic spectra provide a unique application to evaluate the StarNet model, since all details of the spectra are known a priori, and the S/N ratios and wavelength regions can be varied.  Since we will evaluate StarNet on the APOGEE observed data sets, we have applied StarNet to synthetic spectra generated by the APOGEE collaboration.

\subsection{Training and testing StarNet with the APOGEE ASSET spectra}

The synthetic spectra used by the APOGEE consortium have been generated using MARCS and ATLAS9 model atmospheres, as described in \citet{meszaros2012new}, and the radiative transfer code ``ASSET'' \citep{koesterke2008center}. These synthetic spectra were continuum normalized in the same way as the observed APOGEE spectra to facilitate a closer match between the synthetic and observed datasets. Provided in a publicly available 6D spectral grid compressed with Principal Components Analysis (see Table \ref{tab:ASSETgrid} for parameter distribution), we have used these synthetic spectra as a first test of StarNet. To sample a spectrum at any desired location within the synthetic grid, we used a third order interpolation routine between spectra at the existing grid points.

In all instances of training StarNet on synthetic data, we added Gaussian noise to match S/N $\approx$ 20 up to noiseless spectra. Each spectrum was used several times in a particular training process (though only once per training iteration) and therefore different amounts of noise were added with each iteration to mimic the behaviour of multiple visits to a single object in the APOGEE survey. This method also contributes to generalizing the predictive capabilities of StarNet on lower S/N spectra.  

As a first test, we generated a dataset of 300,000 synthetic spectra via random sampling of the stellar parameters within the limits of the ASSET grid. Of these spectra, 260,000 were randomly selected as the reference set. A subset of these (224,000 or 86\% of the reference set) were used to train StarNet. The remaining 36,000 spectra from the reference set were used to cross-validate the model following each training iteration.

\begin{figure}
\centering
\includegraphics[width=\linewidth]{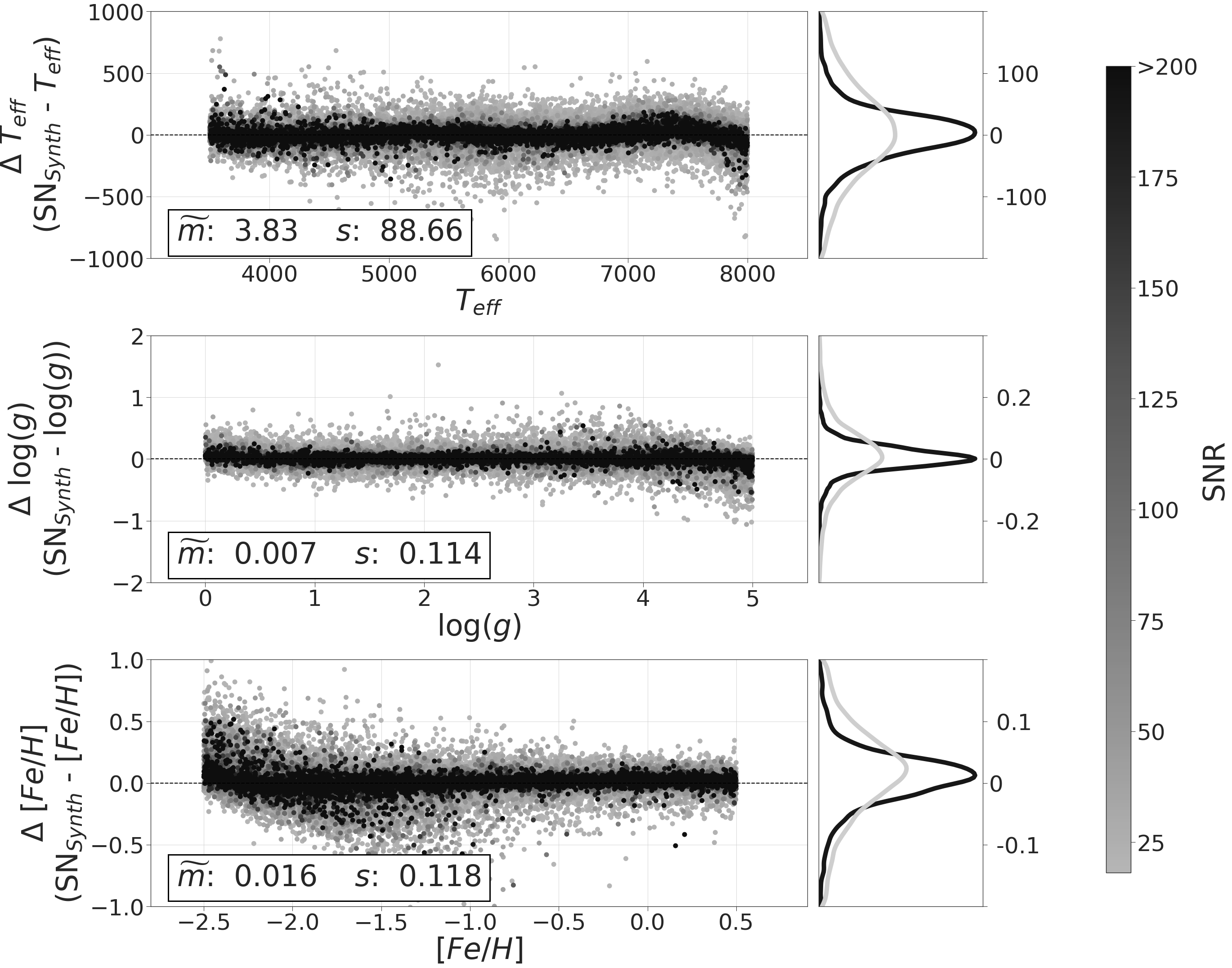}
\caption{StarNet prediction residuals with the generated stellar parameters for a test set of 40,000 ASSET synthetic spectra \citep{koesterke2008center}. StarNet was trained with 224,000 synthetic spectra randomly sampled from the ASSET synthetic grid. Projected residual distributions are shown on the right (black for synthetic spectra with S/N $>$ 80, gray for S/N $<$ 60).  The median value (\~{m}) and standard deviation (s) are calculated in each panel.
\label{fig:trainsynthtestsynth}}
\end{figure}

The 40,000 spectra left out of the reference set were used as a test set, and the residuals between the StarNet predictions and the generated parameters are shown in Fig.~\ref{fig:trainsynthtestsynth}. For all three predicted stellar parameters, the variances of the residual distributions are inversely proportional to the S/N. At [Fe/H]$<$-1.9 and lower S/N, there is a distinct trend where StarNet appears to over-predict the metallicity. In this region, absorption features are not as prominent, therefore the noise effectively causes the StarNet [Fe/H] predictions to be similar for all of these stars (i.e. [Fe/H]$\approx$-2.2).

\subsection {Stellar Parameter Predictions and Precision}
\label{error_propagation}
To evaluate the errors in the StarNet predictions, it is necessary to understand how the output of the model is affected by the uncertainties in the input spectra (\qvec{x}).
Treating the entire network as a vector valued function \qvec{f}(\qvec{x, w, b}) =  
(\qvec{f}$_{\mathrm{T_{eff}}}$, \qvec{f}$_{\mathrm{\logg}}$, \qvec{f}$_{\rm [Fe/H]}$), the Jacobian of the StarNet model for a given star and stellar label, $j$, is given by:

\begin{equation*}
\mathbf{J}(\qvec{x ,w, b}) = \left(\frac{\partial f_j}{\partial x_1}, ... ,\frac{\partial f_j}{\partial x_n}\right)(\qvec{x, w, b})
\end{equation*}

Here we assume that each spectrum is accompanied by a corresponding error spectrum, \qvec{e$_x$}. Propagating the error spectrum with the Jacobian of that spectrum results in an approximation of the statistical uncertainty, {\boldmath$\sigma$}$_{\rm prop}$, of the output prediction for each stellar parameter computed as the product:

\begin{equation*}
\boldsymbol{\sigma}_{\rm prop}^2 \approx \mathbf{J}^2 \cdot {\qvec{e$_x$}}^2
\end{equation*}

Besides restricting our approximation to the small error domain, we do not compute the error from the StarNet CNN model weights for practical reason. To compensate for this deficit,  an empirical intrinsic scatter, {\boldmath$\sigma$}$_{\rm int}$, is derived from the synthetic test data, which is then added in quadrature - element-wise - with the propagated uncertainty.  This gives a total estimated uncertainty:

\begin{equation*}
\boldsymbol{\sigma} = \sqrt{{\boldsymbol{\sigma}}_{\rm prop}^2 + {\boldsymbol{\sigma}}_{\rm int}^2}
\end{equation*}

The intrinsic scatter term is slightly dependent on the region of the parameter-space in which the predicted parameter lies, reflecting the capacity of the physical model to match observed spectral features. For instance, this intrinsic scatter term will be larger for stars that are predicted to have low metallicities compared to those that are found to be metal-rich (see Fig. \ref{fig:trainsynthtestsynth}).  Therefore, we derive this scatter term in bins, within the same range of the ASSET synthetic grid found in Table \ref{tab:ASSETgrid}: for \teff, \logg, and [Fe/H] the bin sizes are 450~K, 0.5~dex, and 0.3~dex, respectively.

\section{StarNet Applications to APOGEE spectra}

The APOGEE survey has been carried out at the 2.5-m Sloan Foundation Telescope in New Mexico, and covers the wavelength range from 1.5 to 1.7 microns in the H band, with spectral resolution R = 22,500. Targets are revisited several times until S/N $\ge$ 100 is reached \citep{Majewski2015}.  All visits are processed with a data reduction pipeline to 1-D and are wavelength calibrated before being combined. The individual visits, unnormalized combined spectra, and normalized combined spectra are all recorded into the publicly-available APOGEE database \citep{nidever2015data}.

The APOGEE Stellar Parameters and Chemical Abundances Pipeline (ASPCAP) is a post data reduction pipeline tool for advanced science products.  This pipeline relies on the nonlinear least squares fitting algorithm FERRE \citep{perez2016aspcap}, which compares an observed spectrum to a grid of synthetic spectra generated from detailed stellar model atmospheres and radiative line transfer calculations. For DR12, this synthetic grid was the ASSET grid used for training in the previous section. The fitting process estimates stellar parameters (\teff, \logg, [Fe/H]) and abundances for 15 different elements (C, N, O, Na, Mg, Al, Si, S, K, Ca, Ti, V, Mn, Fe and Ni). These results are further calibrated with comparisons to stellar parameters from optical spectral analyses of stars from a variety of star clusters \citep{meszaros2015exploring, cunha2015sodium}; see Appendix for more details.

\subsection {Pre-processing of the Input Data}\label{preprocess}

\begin{table*}
\centering
\caption{Cuts applied to APOGEE DR13 for the training and test set}
\label{table:cuts}
\begin{tabular}{cccc}
\hline
\textbf{Cut} & \multicolumn{2}{c}{\textbf{Visit Spectra}} & \textbf{Combined Spectra} \\ \hline
\multicolumn{1}{c|}{No cuts} & \multicolumn{2}{c|}{559484} & 143482 \\
\multicolumn{1}{c|}{\texttt{ASPCAPFLAG}} & \multicolumn{2}{c|}{327594} & 92983 \\
\multicolumn{1}{c|}{\texttt{STARFLAG}} & \multicolumn{2}{c|}{142875} & 45474 \\
\multicolumn{1}{c|}{4000K~\textless ~T$_{\mathrm{eff}}$~\textless ~5500K} & \multicolumn{2}{c|}{127357} & 39017 \\
\multicolumn{1}{c|}{v$_{\mathrm{scatter}}$~\textless ~1km/s} & \multicolumn{2}{c|}{120879} & 37311 \\
\multicolumn{1}{c|}{{[}Fe/H{]} \textgreater ~-3} & \multicolumn{2}{c|}{120879} & 37311 \\
\multicolumn{1}{c|}{log($g$) != -9999} & \multicolumn{2}{c|}{113956} & 35591 \\
\multicolumn{1}{c|}{Combined SNR~\textgreater ~200} & \multicolumn{2}{c|}{53135} & Not applicable \\ \hline
\multicolumn{1}{l|}{\multirow{2}{*}{Partition into reference and test sets}} & \textbf{Training set} & \multicolumn{1}{c|}{\textbf{Cross-Validation Set}} & \textbf{Test set} \\
\multicolumn{1}{l|}{} & 41000 & \multicolumn{1}{c|}{3784} & 21037 \\ \hline
\end{tabular}
\end{table*}

As a test, StarNet was also trained using the APOGEE observed spectra and ASPCAP DR13 stellar parameters \citep[see][]{SDSS2016}. The APOGEE error spectra and the estimated errors from ASPCAP were not included as part of the reference set. This effectively limits the accuracy of StarNet to the reference set parameters. To minimize the propagation of the ASPCAP errors in the StarNet results, we limited the range of our dataset. For example, \citet{holtzman2015abundances} caution against using stars cooler than \teff = 4000\,K where models are less certain, as well as hotter stars where the spectral features are less defined. For these reasons, our dataset was reduced to stars between 4000\,K $\le$ \teff\ $\le$ 5500\,K. 

The APOGEE team have also flagged specific stars and visits for a variety of other reasons. Stars from the DR13 dataset with either a \texttt{STARFLAG} or \texttt{ASPCAPFLAG} were removed from the dataset, which includes stars contaminated with persistence \citep[see][]{jahandar2017, nidever2015data}, stars that were flagged for having a bright neighbour, or having estimated parameters near the parameter grid edge of the synthetic models. Furthermore, stars with [Fe/H] $< -3$ were removed, or those with a radial velocity scatter, $v_{\mathrm{scatter}}$, greater than 1~\kms. Spectra with a high $v_{\mathrm{scatter}}$ are possibly binary stars \citep{nidever2015data}, with potentially duplicate spectral lines.

Implementing these restrictions to the \textit{reference set} allowed the neural network to be trained on a cleaner dataset and learn from more accurate stellar parameters, while placing these restrictions on the \textit{test set} was an attempt to maximize the validity of ASPCAP stellar parameters to be used when comparing with StarNet predictions. All of these cuts to the APOGEE dataset are summarized in Table \ref{table:cuts}.

In addition to the restrictions mentioned above, the reference set only included individual visit spectra from stars where the combined spectrum had S/N $>$ 200. ASPCAP stellar parameter precision is known to degrade at lower S/N \citep{ness2015cannon, casey2016cannon}. However, using high S/N combined spectra as the training input resulted in over-fitting the model (see Appendix \ref{appendix_method}) to spectra with high S/N. To compensate for this issue, StarNet was trained on individual visit spectra, while using the stellar parameters obtained from the combined spectra.  This step allowed the model to be trained on lower S/N spectra with high precision parameters. Limiting our training sample to contain stellar parameters from high S/N spectra was also practical: currently StarNet is not able to weight the input spectra according to the noise properties of the spectra during training. When combining various data sets, which we will address in future work, it might become more critical, but is less relevant for the current analysis.

The last phase of the data pre-processing was normalizing the spectra. Both the Cannon~2 \citep{casey2016cannon} and ASPCAP have implemented independent continuum normalization techniques; in addition, the APOGEE DR13 spectra were further normalized in certain spectral regions. The ASSET synthetic spectra were similarly normalized to facilitate proper matching with the data. It has been suggested \citep[e.g.,][]{casey2016cannon} that the ASPCAP continuum normalization techniques result in a S/N dependency. This is a potential limitation of StarNet when it has been trained on synthetic spectra, since these spectra were ASPCAP normalized without the addition of noise beforehand. Any non-linearities in continuum normalization would be excluded from the synthetic spectra dataset. If this inconsistent normalization is indeed a problem, it could create an inherent mismatch between a synthetic spectrum and a low S/N APOGEE spectrum of identical stellar parameters, leading to erroneous stellar parameter estimates.

This potential for a continuum normalization bias motivated us to adopt a simple and linear normalization method in StarNet; for training and testing on APOGEE spectra, the spectra were split into blue, green, and red chips, and each chip was divided by its median flux value. The three chips were then combined to create a single spectrum vector. Given that normalization procedures usually do not need external information, our simple approach is also testing the network capabilities compared to a more physically-driven continuum removal. Further analysis is, however, required to determine the full impact of continuum normalization in this NN approach.

\subsection {Training and testing StarNet with APOGEE spectra}\label{test_real_on_real}

StarNet has been trained and tested on the ASPCAP stellar parameters (\teff, \logg, [Fe/H]) corresponding to individual visit spectra and combined spectra, respectively. As discussed previously, this included stars from the APOGEE DR13 dataset with the cuts outlined in Table \ref{table:cuts}. There were 17,149 stars with S/N $>$ 200 that met these requirements, of which 2651 were used as part of the test set, and 14,498 stars (containing 44,784 individual visits) were used as the StarNet reference set.  This is slightly less than 10\% of the total APOGEE DR13 dataset. A subset of these individual visits (41,000 or 92\% of the reference set) were randomly selected for the training set. The remaining 3,784 visit spectra were used to cross-validate the model StarNet following each forward propagation during training.  No significant deviations were found between random selections for training and cross-validation samples.

\begin{figure}
\centering
\includegraphics[width=\linewidth]{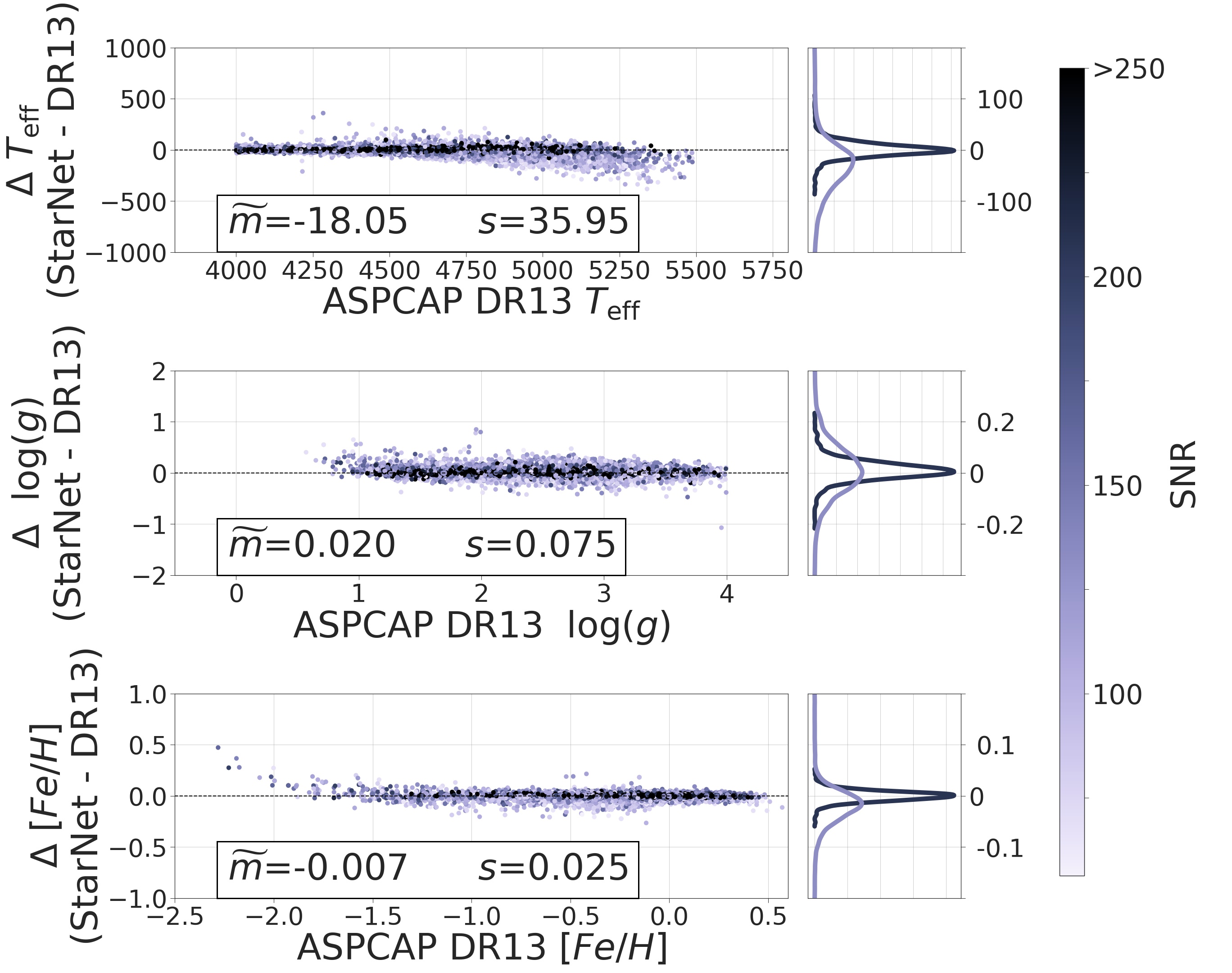}
\caption{Residuals of StarNet predictions and ASPCAP parameters for a test set of 21,037 combined spectra over a large range of S/N. StarNet was trained on 41,000 individual visit spectra from the APOGEE DR13 dataset. As the S/N decreases, small deviations are seen for the hotter stars, stars at the lower end of the surface gravity range, and the most metal-poor stars, which are likely due to a sample size bias in the training set. Projected residual distributions are shown on the right (dark purple for spectra with S/N $>$ 200, light purple for S/N $<$ 100). The median value (\~{m}) and standard deviation (s) are calculated in each panel, as in Fig.~\ref{fig:trainsynthtestsynth}.}
\label{HIGHSNR}
\end{figure}

Once trained, StarNet was applied to the test set containing both high and low S/N combined spectra. The StarNet predictions are compared to ASPCAP parameters in Fig.~\ref{HIGHSNR}. For the high S/N spectra, StarNet predictions show excellent agreement with the ASPCAP DR13 results. 
For the lower S/N spectra, there are more and larger deviations between StarNet and DR13. For example, at high temperatures (\teff $>$ 5000\,K), StarNet predicts lower effective temperatures than DR13.   The quality of the model predictions depends on the number of stars in the training set that span the parameter space of the test set, therefore we suggest that these deviations are due to an insufficient number of stars with \teff $>$ 5000\,K ($\sim$4\%) in our reference set.  Similarly, there are few stars at low metallicities, [Fe/H] $< -1.5$ ($\sim$0.2\% of the reference set), and therefore applying StarNet to the most metal-poor stars also results in larger deviations from the DR13 results. Sample size bias, as seen in these two regions, reinforces the need for large training sets for deep neural networks such as StarNet.

Another potential source of error in the predictions at high temperatures and low metallicities lies in the spectra themselves. In these regions, the spectral features are weaker, therefore the neural network will struggle to locate the most important features during training.  Similarly, the parameters determined by ASPCAP will have larger intrinsic uncertainties. Furthermore, these effects can be amplified when testing on lower S/N spectra.

After training, StarNet was then applied to 148,724 stars in the APOGEE DR13 database. This final inference step is very fast: taking about 1 minute of CPU time for the whole data set. The short amount of time required to make these predictions is a huge advantage for the rapid data reduction of large spectral surveys.

\begin{figure*}
\centering
\begin{multicols}{2}
\includegraphics[width=\linewidth]{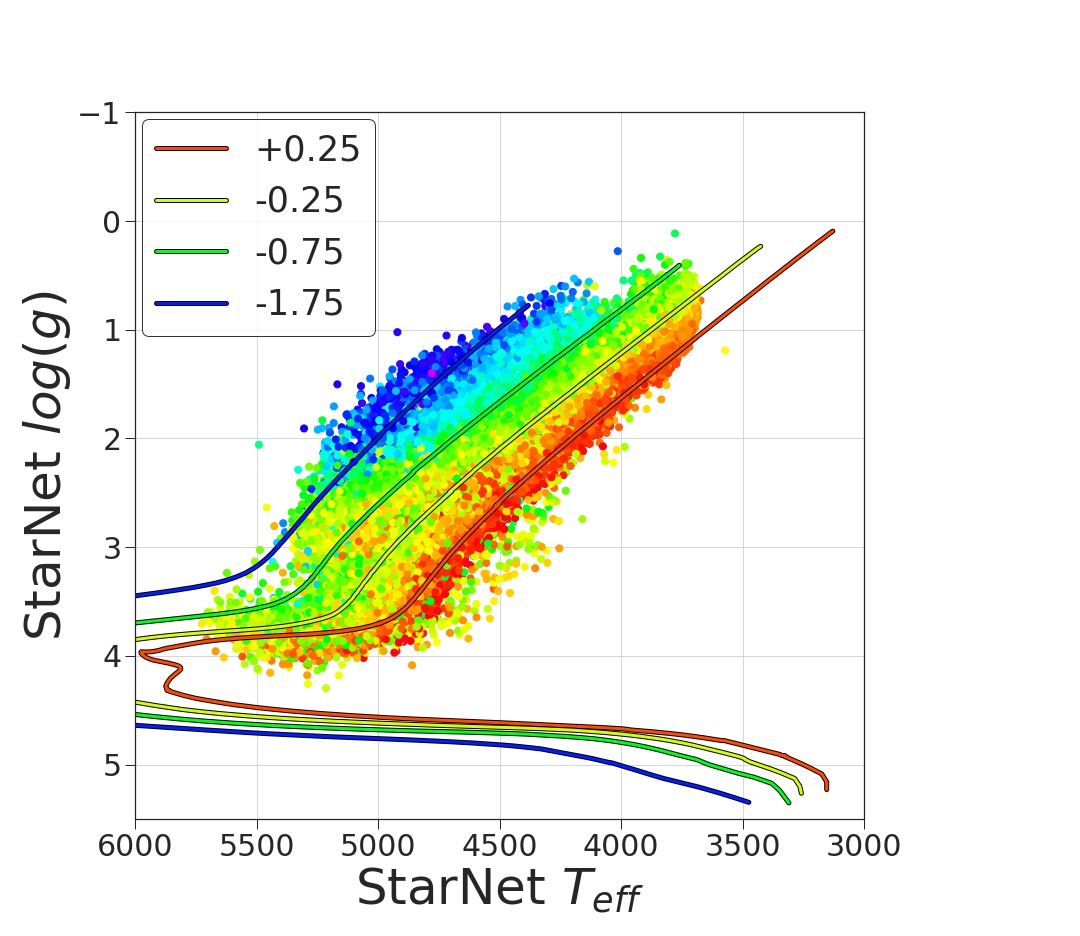}\par
\includegraphics[width=\linewidth]{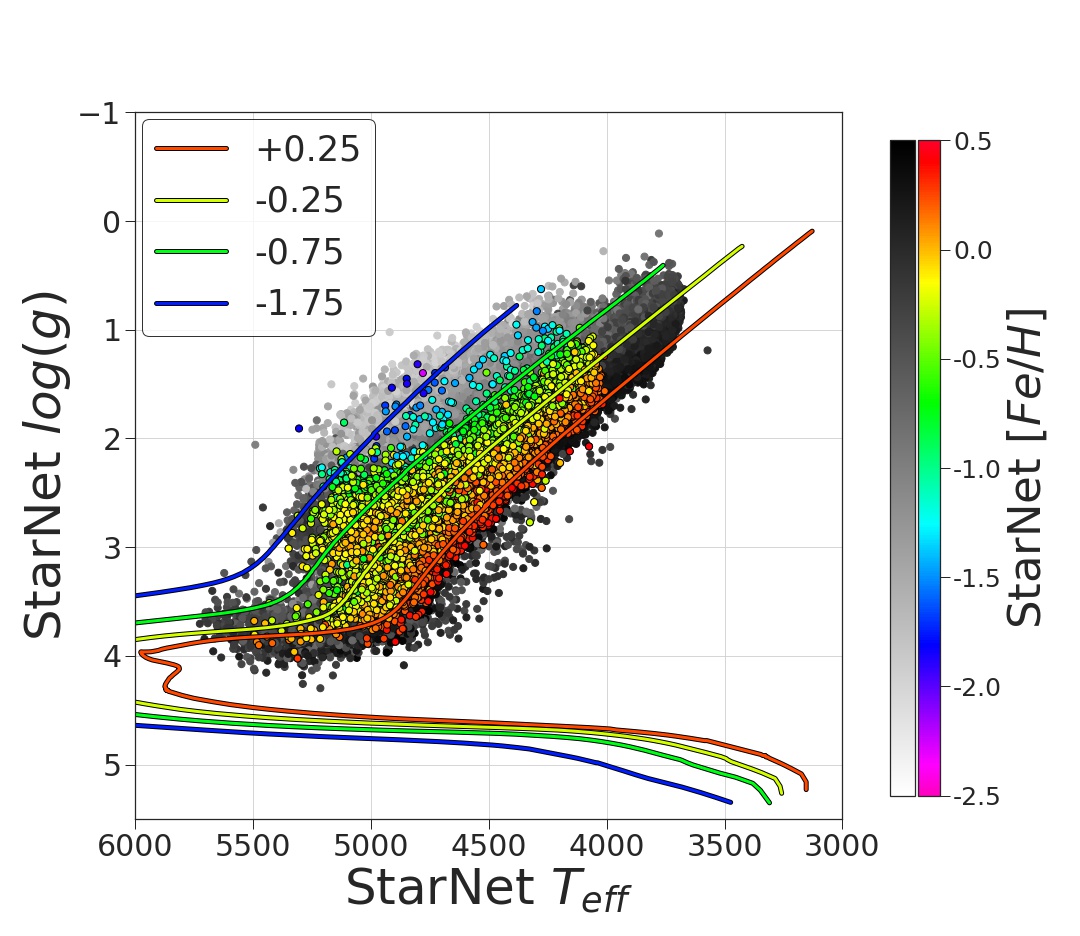}\par
\end{multicols}
\caption{Left panel:  stellar parameters from StarNet for 99,211 stars, showing {\logg} parameters against {\teff} across a wide range of [Fe/H] (see colour bar).  Right panel:  the StarNet reference set of 14,498 stars highlighted over top of the full sample of 99,211 stars (in gray).   Stellar isochrones are shown in both panels \citep[from][]{dotter2008dartmouth}, with ages of 5 Gyr and [Fe/H] values found in the upper left corner (corresponding to the colour bar).
\label{veracity}}
\end{figure*}

StarNet predictions for 99,211 spectra are shown in Fig.~\ref{veracity}, where stars with ASPCAP \teff, \logg, or [Fe/H] values of -9999 were removed along with those that are presumed to be M dwarf stars (see Section \ref{m_dwarfs}). StarNet predictions are compared to stellar isochrones to show the expected trend in the parameter relationships; four isochrones with different metallicities ([Fe/H] = 0.25,$-$0.25,$-0.75$ and $-$1.75) were generated from the Dartmouth Stellar Evolution Database \citep{dotter2008dartmouth}, with an age of 5 Gyr and [$\alpha$/Fe]=0.  Fig.~\ref{veracity} also shows the StarNet reference set of 14,498 stars, highlighted to show the spread in their abundances across the parameter space.

\begin{figure*}
\centering
\begin{multicols}{2}
\includegraphics[width=\linewidth]{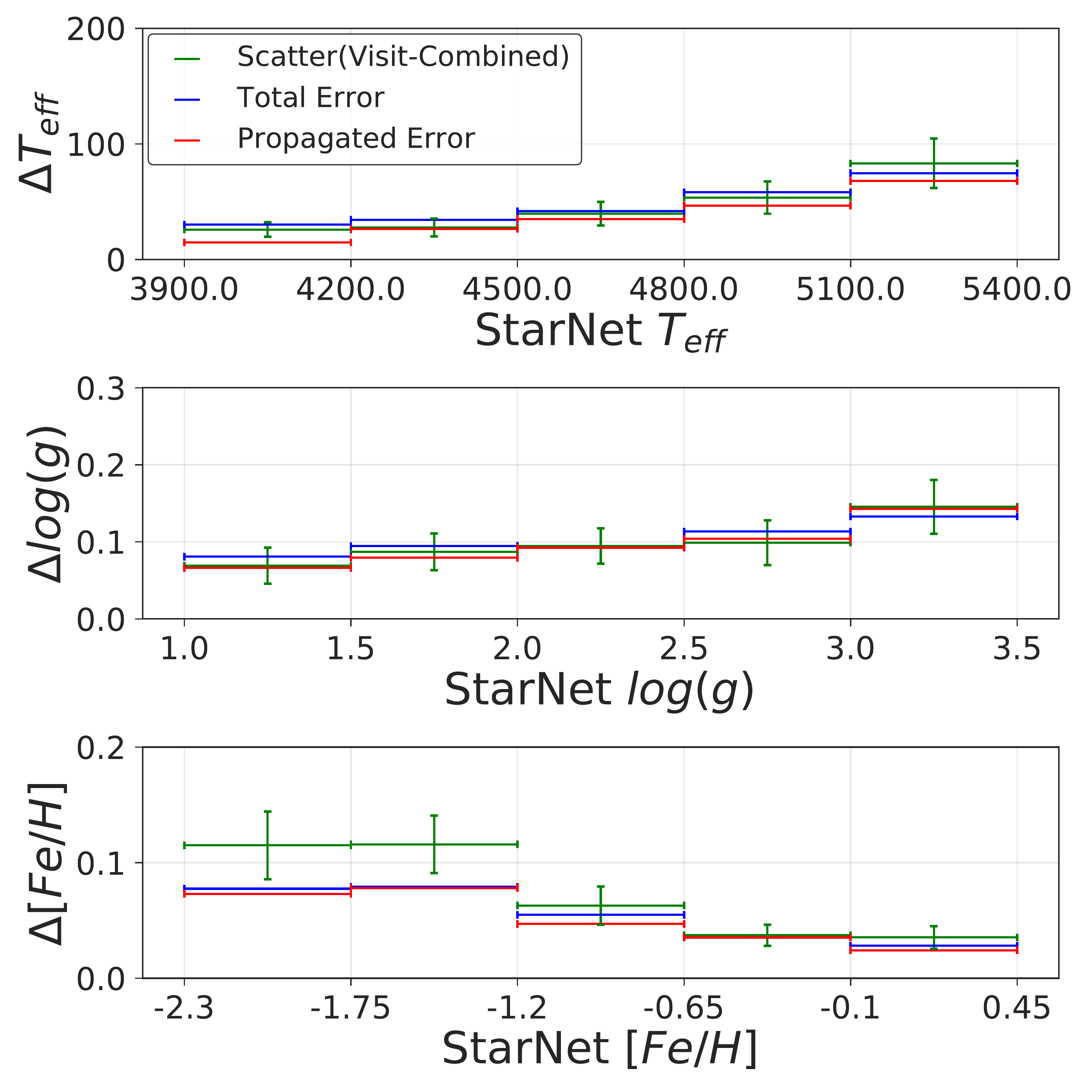}\par
\includegraphics[width=\linewidth]{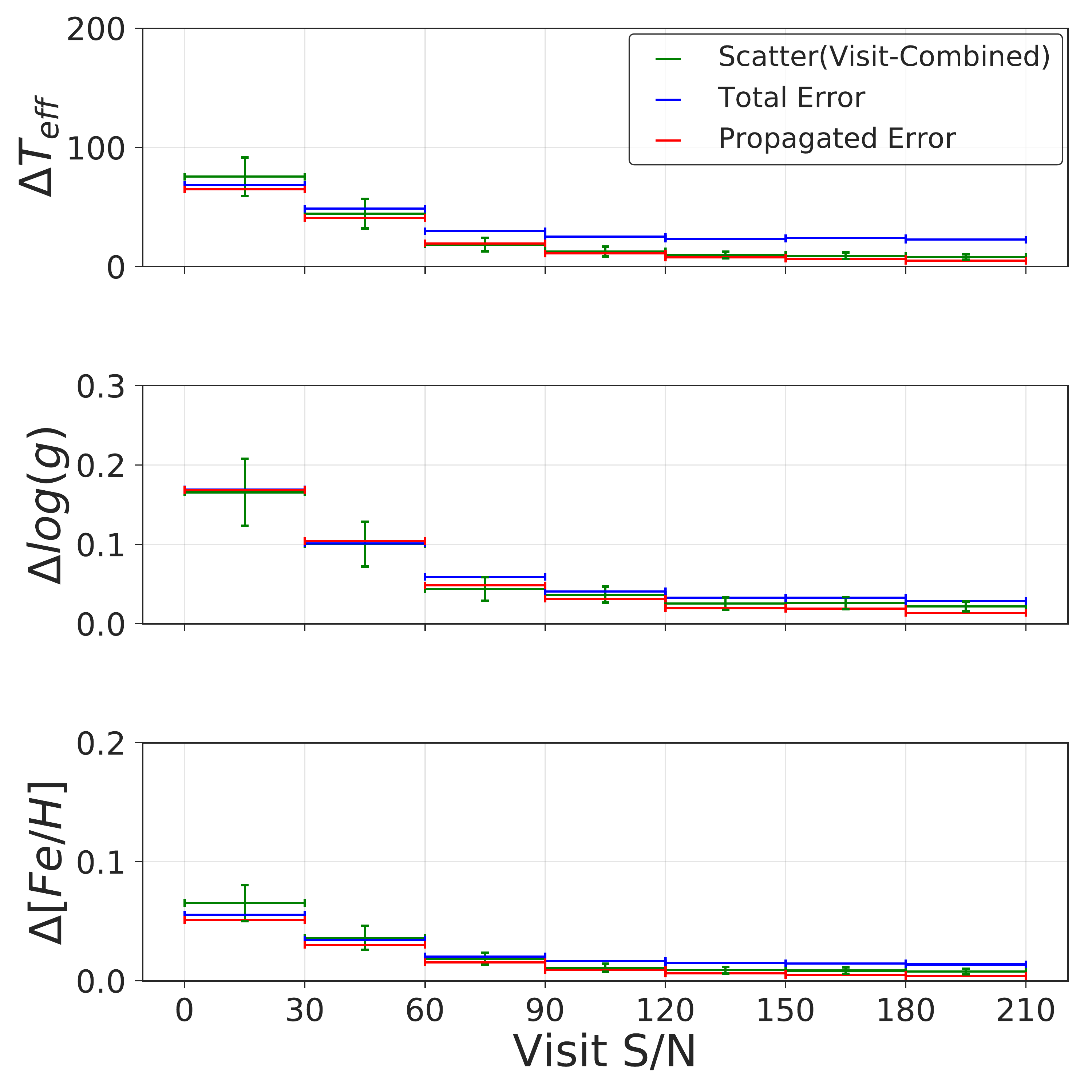}\par
\end{multicols}
\caption{Comparison of StarNet predictions on individual visits to predictions for the corresponding combined spectra. Left panel: the comparisons are made to the parameter-space, where each parameter bin includes 100 randomly selected spectra.  The largest deviations are found at lower metallicities and higher temperatures.  Right panel: comparisons are made based on the S/N of the individual visits, where each S/N bin contained 230 spectra. In both plots the ``Scatter (Visit-Combined)'' is the standard deviation in the residuals between the predictions for the individual visits and combined spectra for objects with more than 4 individual visits. The ``Propagated Errors'' are the error terms due to the error spectrum, whereas the ``Total Errors'' are the propagated error and ``Intrinsic Scatter'' terms (see Section \ref{error_propagation}) added in quadrature. StarNet was trained on 41,000 individual visit spectra from the ASPCAP DR13 dataset.}
\label{vists_vs_comb}
\end{figure*}

To analyze the StarNet model self-consistency, predictions were made for both the individual visit spectra and combined spectra for the same objects. Targets with four or more individual visits were selected for this test, and the differences in the residuals were calculated. 
For each stellar parameter bin, the mean value for 100 objects is shown in Fig.~\ref{vists_vs_comb} (left panel), whereas there were 230 objects in each S/N bin (right panel). The propagated and total errors shown are discussed in Section \ref{error_propagation}.
When comparing StarNet predictions for the individual visits to those for the combined spectra, the results are quite consistent across the majority of the parameter-space, though as discussed previously, there is a noticeable increase in the deviations at lower metallicities ([Fe/H] $<$ -1.2 dex) and higher temperatures ({\teff} $>$ 5100K).  
Also as expected, the scatter increases at lower S/N, although only marginally until S/N $\lesssim$ 60. Even at the lowest values (S/N $\sim$ 15) the results are quite consistent.  The ability to predict well on lower S/N spectra is largely due to the fact that StarNet was trained on individual visits - rather than combined spectra only - with high validity stellar parameters.

The trends seen in the residuals between predictions for individual visits and combined spectra (i.e., an increase in scatter at lower metallicities and lower S/N) are also reflected in the StarNet propagated errors (see Section \ref{error_propagation}), also shown in Fig. \ref{vists_vs_comb}. This provides confidence that the error propagation methods used give an adequate estimate of the uncertainties in the StarNet predictions.

\subsection{Model Selection}
\label{model_comparison}

\begin{figure}
\centering
\includegraphics[width=\linewidth]{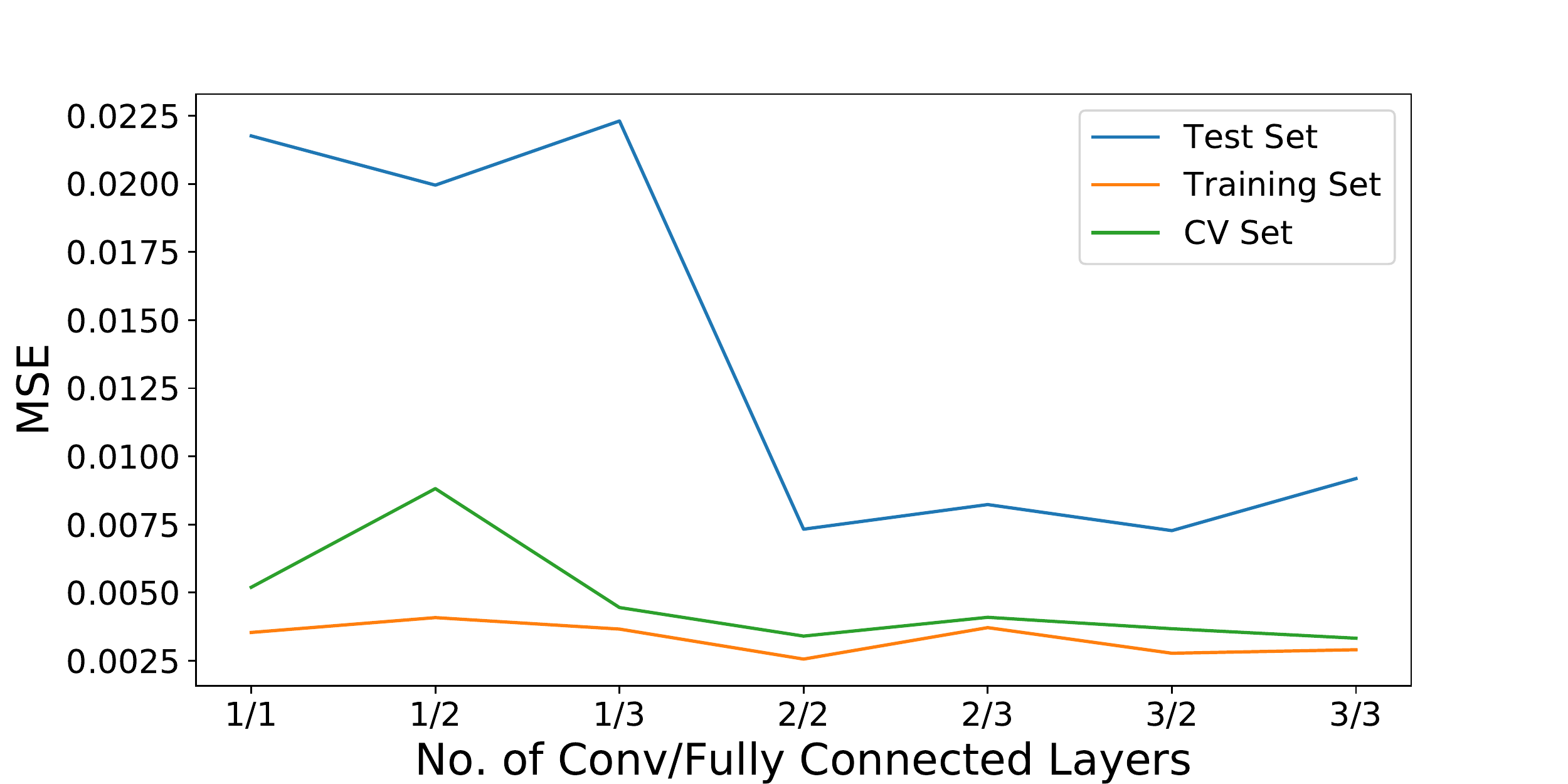}
\caption{The Mean Squared Error (MSE) between the normalized target parameters and StarNet predictions are plotted against the different combinations of convolutional and fully connected layers. These results were used to help select the model architecture of StarNet.}
\label{model_selection}
\end{figure}

Deep learning architectures, such as StarNet, typically involve experimentation and tuning of hyper-parameters in order to converge to an adequate model. Some of these hyper-parameters include: the number of filters in each convolutional layer; the length of the filters that are convolved across the inputs in each of these convolutional layers; the number of connection nodes in each fully connected layer; the pooling window size in the maxpooling layer (the number of inputs which are compared against each other to find the maximum value); and the learning rate of the gradient descent optimizer.
The final selection of hyper-parameters was automated as follows: 
to select the optimal model architecture, a hyper-parameter optimization was run in two stages. During the first stage, the number of convolutional layers, fully connected layers, filters, and nodes were varied randomly, along with the length of the filters, the pooling window size, and the learning rate. This allowed the optimizer to test different combinations of the number of convolutional and fully-connected layers with a variety of other hyper-parameters to ensure that each combination could reach its maximum predictive potential. The same training and cross-validation sets (as described in Section \ref{preprocess}) were used during the training of these models, and the same test set (discussed in Section \ref{test_real_on_real}) was used to test each model. The metric used for model comparisons was the ``Mean Squared Error" (MSE) between the target and predicted parameters. To have each parameter weighted equally, the parameters were normalized to have approximately zero-mean and unit variance. An example of the best performing models, evaluated for each combination of layers, is shown in Fig.~\ref{model_selection}.

It was found that increasing the number of layers improved the model predictions, but that there was a plateau in performance when combining two convolutional layers with two fully connected layers. The addition of more layers onto this architecture does not clearly improve the prediction results, while fewer layers led to significantly worse MSE.

A second hyper-parameter optimization was then started, where the number of convolutional and fully-connected layers were fixed at two each, while the remaining hyper-parameters were varied and optimized using a Tree-structured Parzen Estimator \citep[TPE,][]{NIPS2011_4443}.
The model selected by this second stage of hyper-parameter optimization was similar to that shown in Fig.~\ref{fig:StarNet}, and this model was used as the starting point for our model architecture selection. Small changes were made to discover possible improvements until the StarNet model architecture performed consistently with adequate results.

While simpler models could perform comparably well to the StarNet architecture when trained on synthetic data, it was decided that a single model architecture should be used to train on both synthetic data and APOGEE spectra.

Earlier studies found that applying neural networks to stellar spectra only necessitates at most two hidden layers with fewer nodes than the amount used in StarNet \citep{bailer2000stellar}, but these were applied to much smaller data sets. Current machine learning methods - along with large datasets - allow for more complex models to improve performance while still avoiding over-fitting. The max pooling layer reduces the degrees of freedom of the model by minimizing the impact of unnecessary weights and therefore simplifying the model function. To speed up convergence for the StarNet model fit, we use the \textit{He Normal} \citep{he2015init} weight initialization, also permitting for convergence of deeper models and eventually enabling the neural network to find more complex features. In short, the \textit{He Normal} initializer, ReLU activation, and ADAM optimizer (see Appendix \ref{appendix_method}) allow for a deeper model to reach convergence more easily and at an accelerated rate, while the max pooling layer simplifies the model rank complexity. The use of a cross-validation set is another technique implemented in StarNet to mitigate over-fitting. These techniques and methods are described further in Appendix \ref{appendix_method}.

\subsection{Comparisons with The Cannon}

The Cannon \citep{ness2015cannon} is a data-driven, generative model that shares the same limitations as a supervised learning approach, i.e., both rely on a reference data set. Unlike StarNet, the generative approach uses stellar parameters as the inputs and spectra as the outputs during training. The Cannon uses a quadratic polynomial function to translate the stellar parameters into spectra, and the best-fitting coefficients of the function are found through least-squares fitting. During the test phase, the stellar parameters are determined with another regression step, where the stellar parameters are varied until they produce a spectrum that best matches the observed spectrum.

\begin{figure*}
\centering
\begin{multicols}{3}
\includegraphics[width=\linewidth]{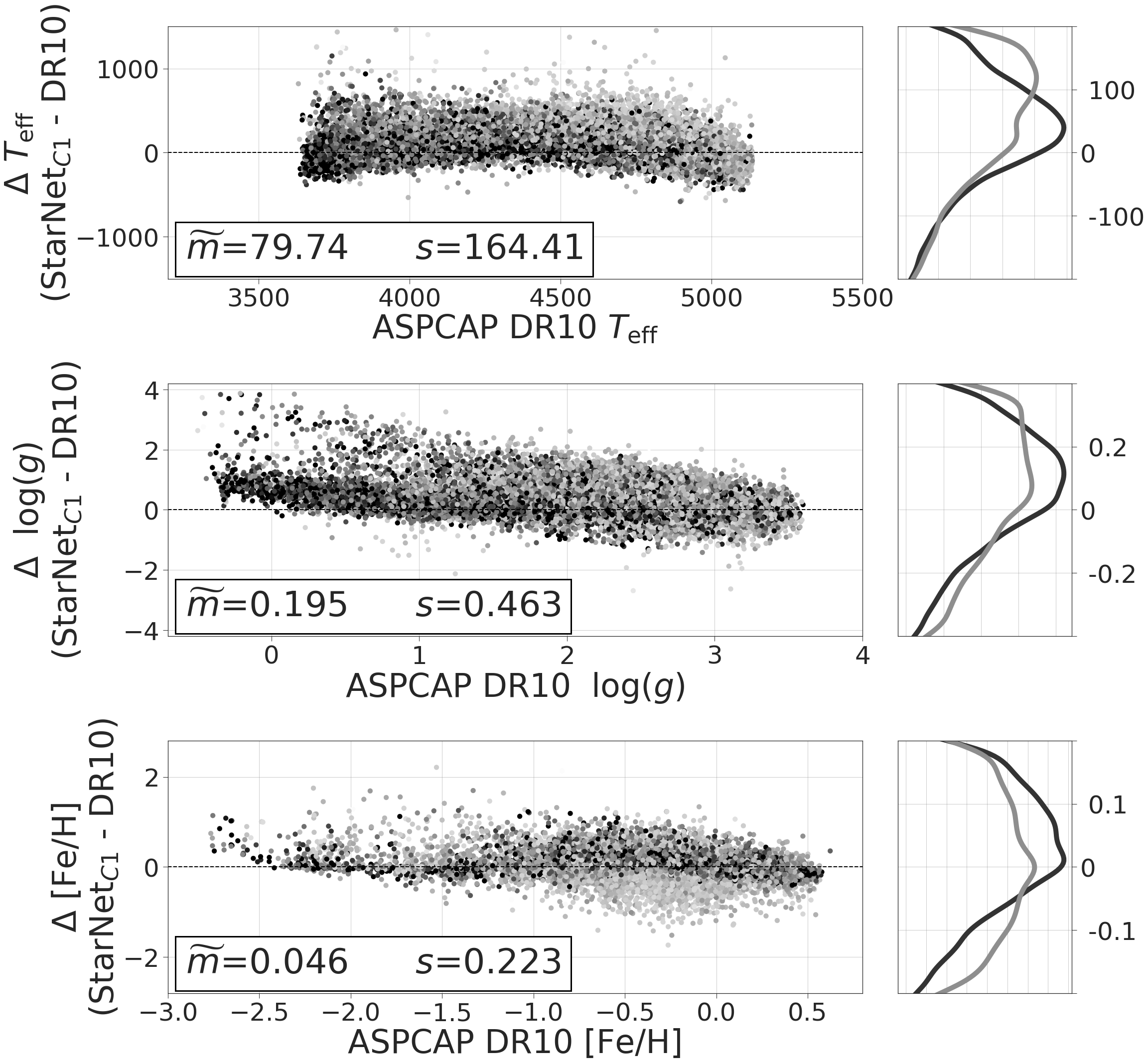}\par
\includegraphics[width=\linewidth]{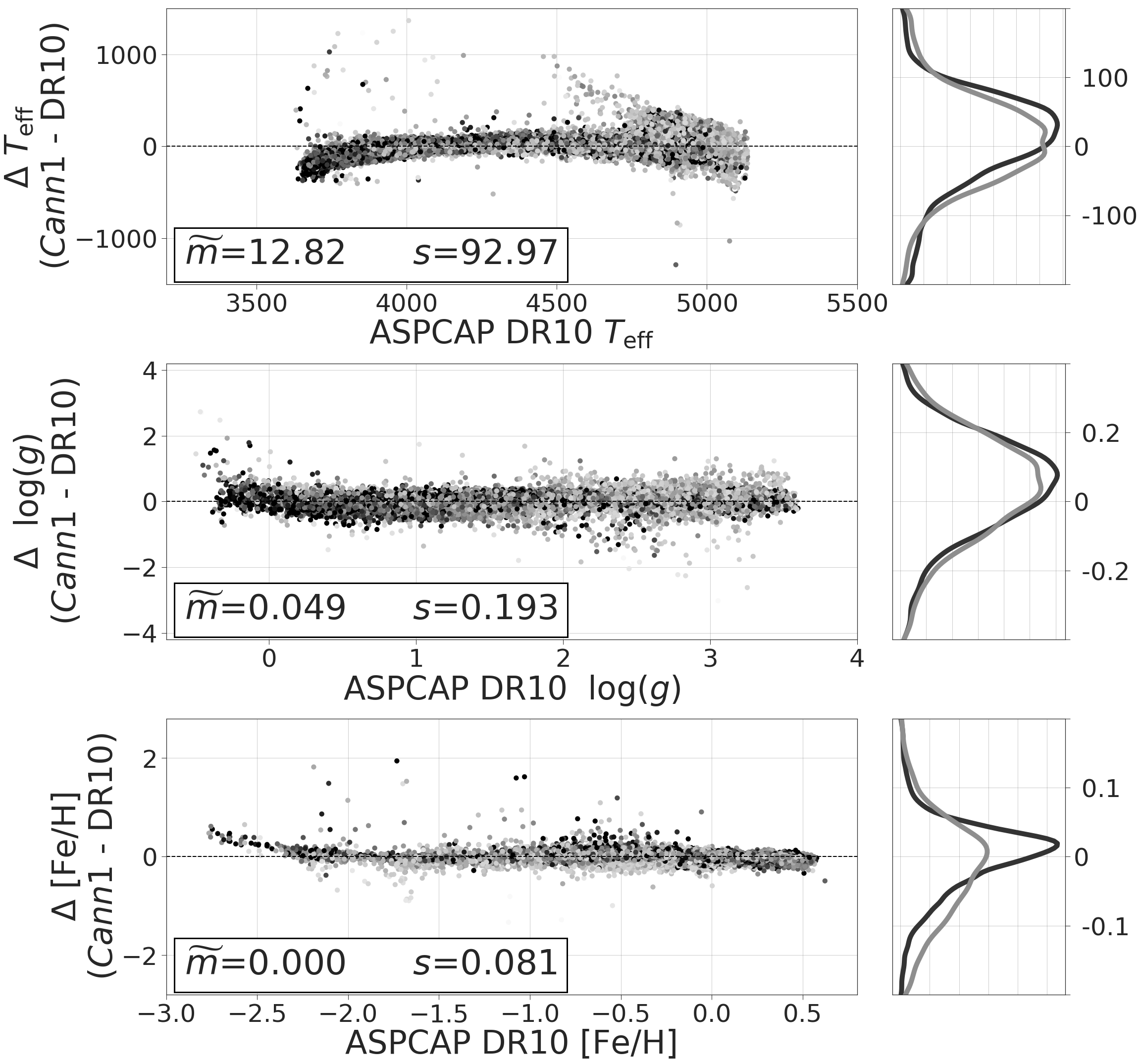}\par
\includegraphics[width=\linewidth]{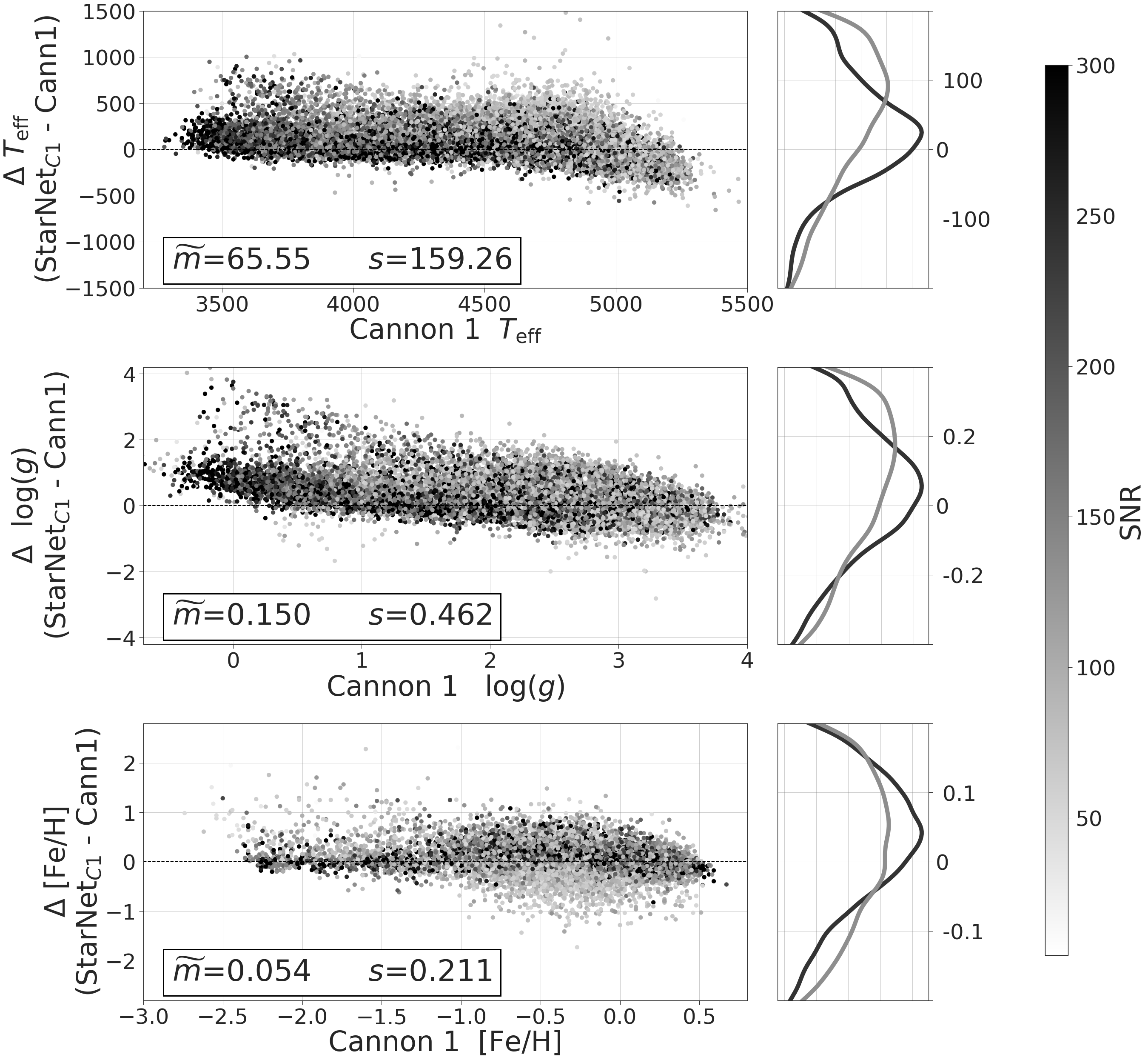}\par
\end{multicols}
\caption{
Comparison of StarNet$_{C1}$ results with ASPCAP (left panel) and  StarNet$_{C1}$ results with The Cannon 1 (right panel), as well as comparisons between The Cannon 1 and ASPCAP (center panel). StarNet$_{C1}$ was trained on APOGEE DR10 combined spectra from the same 542 stars that The Cannon 1 used for training. The test set includes combined DR10 spectra that had both ASPCAP and The Cannon 1 predictions. The median value (\~{m}) and standard deviation (s) are calculated in each panel, as in Fig.~\ref{fig:trainsynthtestsynth}.  (Note that the range in the plot axes differ from the other plots in this paper). 
\label{cannon1_test}}
\end{figure*}

The Cannon was originally trained on APOGEE DR10 \citep{meszaros2013calibrations} combined spectra from 542 stars in 19 clusters. These are far fewer training examples than StarNet uses from DR13, and yet The Cannon was able to predict stellar parameters for the DR10 spectra extremely well. In an effort to compare the two techniques, we trained the same StarNet architecture, labeled StarNet$_{C1}$, on the same data set of 542 stars from DR10.   StarNet$_{C1}$ was then applied to 29,891 combined spectra that had both ASPCAP and The Cannon predictions for stellar parameters. These are compared in Fig.~\ref{cannon1_test}, where the limitations of the StarNet$_{C1}$ model are clearly seen.
The machine learning method used for StarNet requires a {\it large} training sample that spans a wide parameter space. Training on 542 combined spectra does not span enough of the parameter range at an accumulated S/N to fit the complex StarNet convolutional neural network.  This can also be seen in the large residuals when comparing StarNet$_{C1}$ predictions and ASPCAP DR10 parameters. Predictions for low S/N spectra show the most significant scatter.

The requirement of large training sets that adequately represent the test set is a limitation of many deep learning methods, such as StarNet. However, as larger data sets become available, then these methods can be extremely effective, as discussed in the next Section.

\subsection {Comparisons with The Cannon 2}

\begin{figure*}
\centering
\begin{multicols}{3}
\includegraphics[width=\linewidth]{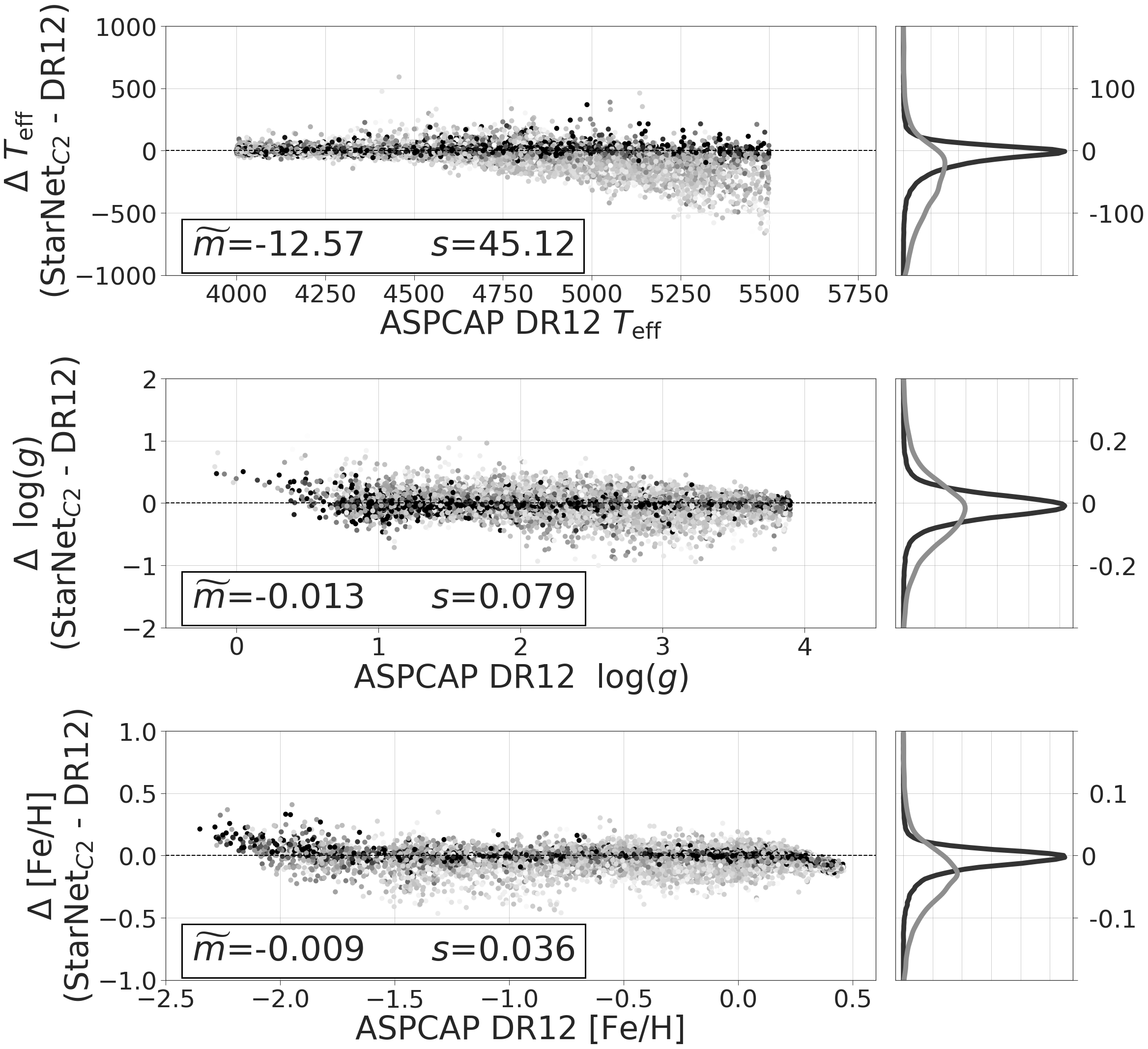}\par
\includegraphics[width=\linewidth]{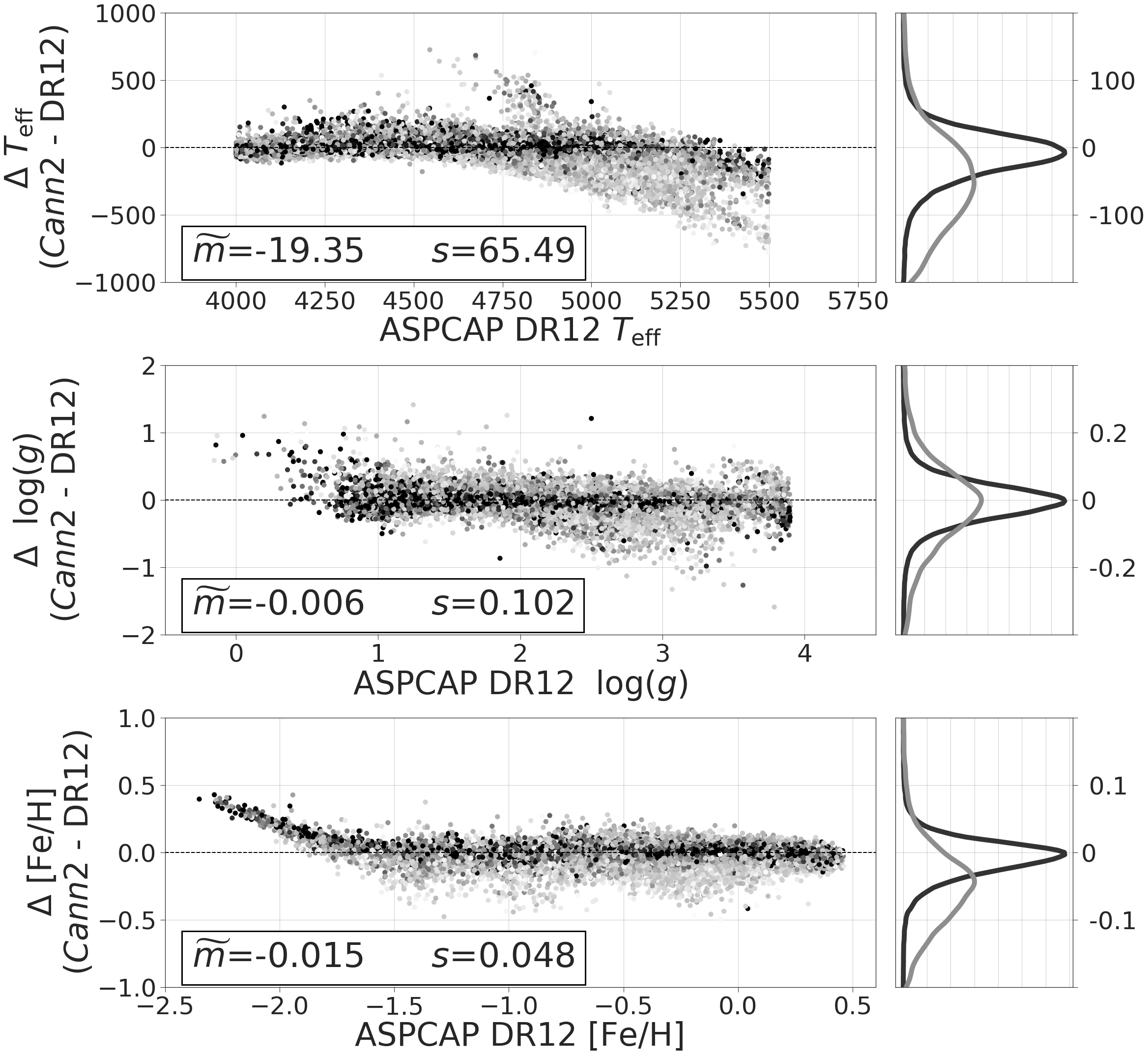}\par
\includegraphics[width=\linewidth]{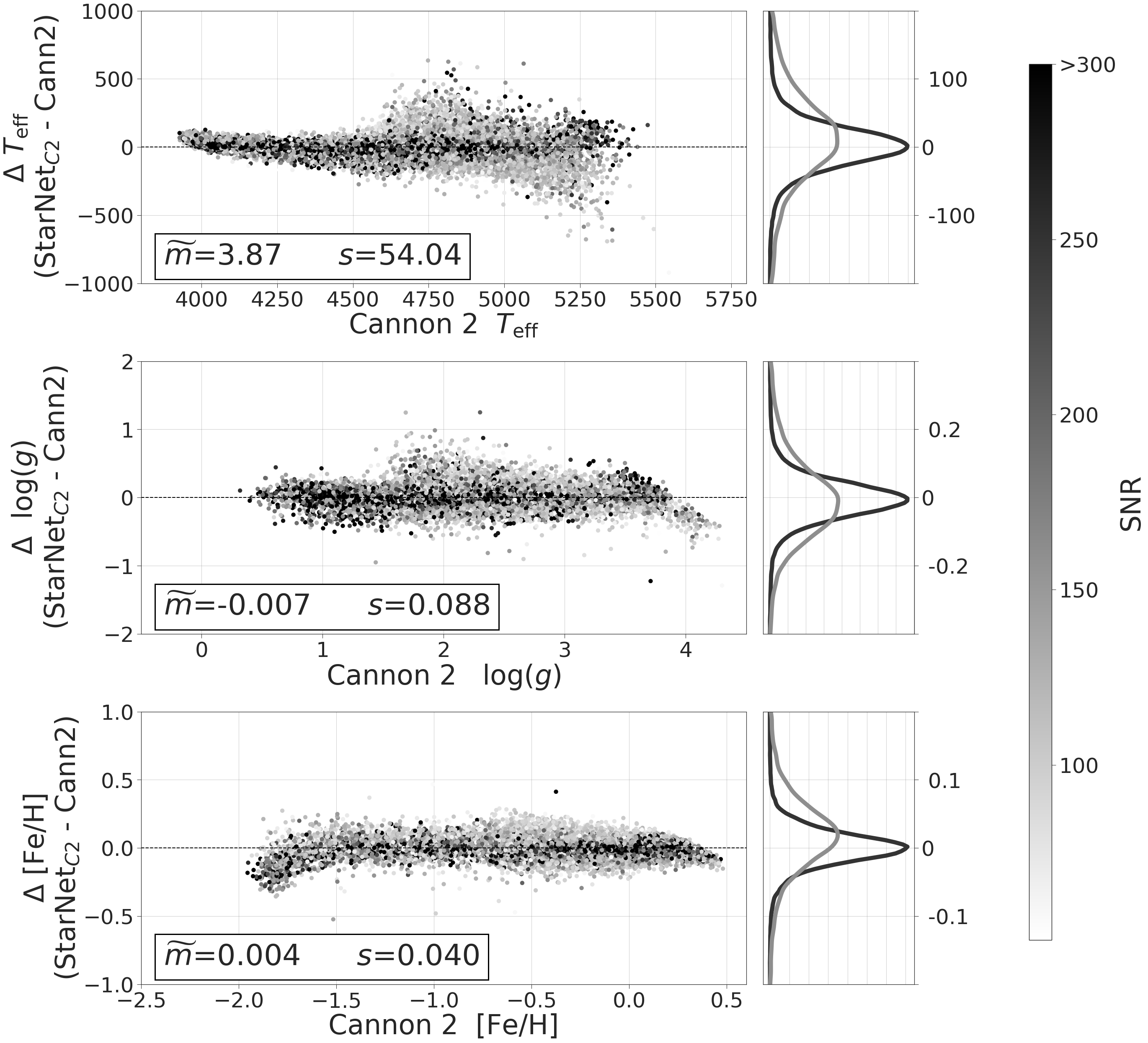}\par
\end{multicols}
\caption{
Comparison of StarNet$_{C2}$ results with ASPCAP (left panel) and  StarNet$_{C2}$ results with The Cannon 2 predictions (right panel), as well as comparisons between The Cannon 2 and ASPCAP (center panel).  All parameters are based on the DR12 data set.  StarNet was trained on individual visits from similar stars as those used for training The Cannon 2. The test set used to compare the two methods was also similar to that used by The Cannon 2.  The median value (\~{m}) and standard deviation (s) are calculated in each panel, as in Fig.~\ref{fig:trainsynthtestsynth}.
}
\label{cannon2_test}
\end{figure*}

The Cannon 2 \citep{casey2016cannon} is a continuation of The Cannon project. In addition to the three stellar parameters predicted by The Cannon, The Cannon 2 predicts 14 additional chemical abundances. The Cannon 2 is also trained on individual visit spectra, but it trains on a larger number of stars than The Cannon.  Also, The Cannon 2 adds a regularization term in the function at training, to reduce the likelihood of the model over-fitting the training data.

The training set for The Cannon 2 consisted of stars with $200 <$ S/N $< 300$, [Fe/H] $> -3$, [X/Fe] $< 2$, [$\alpha$/Fe] $> -0.1$, [V/Fe] $> -0.6$, $V_{\mathrm{scatter}}<$ 1 km\,s$^{\rm -1}$, and excluded targets flagged with the \texttt{ASPCAPFLAG}. These cuts removed stars near the edge of the grid of synthetic models, along with other sources of error resulting in poor ASPCAP parameter determinations.  However, it did not take into account spectra flagged with persistence. The Cannon 2 training set consisted of visits from 12,681 stars.

To make an adequate comparison between the two models, StarNet$_{C2}$ was trained on 39,098 visits from 12,681 stars that met the same restrictions set by The Cannon 2 training set, from APOGEE DR12 \citep{holtzman2015abundances}.  A test set of 85,341 combined spectra with 4000\,K $<$ \teff $<$ 5500\,K and {\logg} $<$ 3.9 was adopted, with similar restrictions as The Cannon 2 test set.

We compile the Root Mean Square Error (RMSE) and the Mean Absolute Error (MAE) for an identical test set for both The Cannon 2 and StarNet$_{C2}$ in Table \ref{cannon2_table}.  This confirms the visual inspection of the residuals in Fig.~\ref{cannon2_test}, that StarNet$_{C2}$ is capable of predicting values closer to the ASPCAP stellar parameters. 
Also seen in Fig.~\ref{cannon2_test}, StarNet$_{C2}$ differs from ASCAP at lower metallicities and hotter temperatures, when the S/N of the spectra is lower. These effects are also seen in The Cannon 2 predictions. In fact, a comparison of StarNet$_{C2}$ to The Cannon 2 shows that the effect is larger in The Cannon 2 results. 

\begin{table}
\centering
\caption{Comparison of the StarNet$_{C2}$ and The Cannon 2 results for a test set of 85,341 combined spectra from APOGEE DR12. Metrics used are the mean absolute error (MAE), and root mean squared error (RMSE), with respect to the same stars.}
\label{cannon2_table}
\begin{tabular}{llll}
\hline
Metric                & Parameter  & StarNet$_{C2}$ & Cannon 2 \\ \hline
\multirow{3}{*}{MAE}  & \teff      & 31.2    & 46.8     \\
                      & \logg      & 0.053   & 0.066    \\
                      & {[}Fe/H{]} & 0.025   & 0.036    \\ \hline
\multirow{3}{*}{RMSE} & \teff      & 51.2    & 71.6     \\
                      & \logg      & 0.081   & 0.102    \\
                      & {[}Fe/H{]} & 0.040   & 0.053    \\ \hline
\end{tabular}
\end{table}

\subsection{Training StarNet with Synthetic Spectra}

A crucial next step in the development of a NN approach to analyzing stellar spectra is to show that StarNet can also predict stellar parameters without an external pipeline.   In this Section, we present stellar parameter results for APOGEE data after training StarNet on synthetic spectra only.   Our goal is to show that StarNet can operate as a standalone data processing pipeline, producing an independent database of stellar parameters that does not depend on previous ASPCAP pipeline results for training.

\subsubsection{Synthetic Gap}

Differences in feature distributions between synthetic and observed spectra is referred to as the \textit{synthetic gap}. To probe the feasibility of training StarNet on synthetic spectra and accurately predict the parameters from observed spectra, it was necessary to ensure that the synthetic gap was relatively small. Since each spectrum consists of 7214 data points, or represents a point in 7214-dimensional space, one would expect a synthetic and observed spectrum with the same parameters to occupy approximately the same region in this space if the gap was indeed small. Visualizing this space is possible only through dimensionality reduction.  We employed the t-Distributed Stochastic Neighbour Embedding \citep[t-SNE,][]{maaten2008visualizing}, a technique often used in machine learning to find clusters of similar data in a two dimensional space. This two dimensional space is comprised of arbitrary variables - not physical stellar parameters - meant to serve as a lower dimensional representation of the spectra to facilitate visualization.

\begin{figure}
\centering
\includegraphics[width=\linewidth]{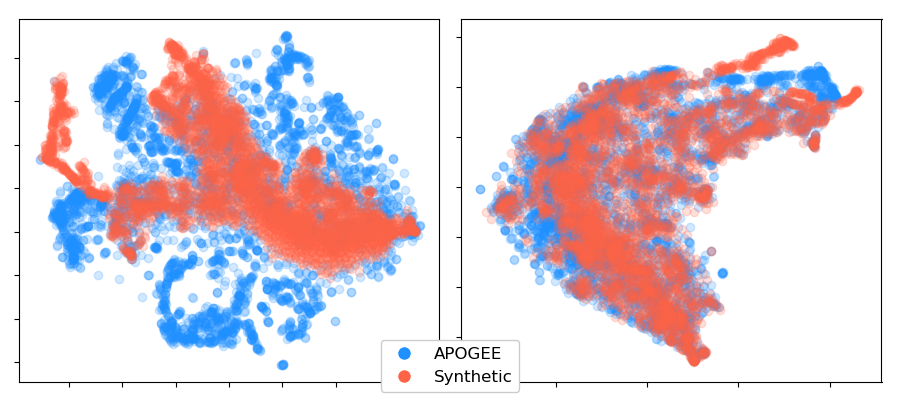}
\caption{t-SNE visualization of the synthetic and APOGEE spectra before zero-flux substitution through nearest neighbour interpolation (left) and after (right).
\label{fig:tsne}}
\end{figure}

A subset of 4000 APOGEE DR13 spectra with S/N$>$200 and known ASPCAP parameters were randomly selected, along with 4000 interpolated synthetic ASSET spectra with the same parameters. After applying t-SNE, a distinct separation could be seen between the synthetic and APOGEE spectra (left image in Fig. \ref{fig:tsne}).  Examination of mismatched spectra showed that there were zero-flux values in wavelength bins along the spectra, a method used in the APOGEE pipeline to flag bad pixels. Unfortunately, it is not possible to mask these values at the testing stage with the current implementation of StarNet, thus a nearest-neighbour interpolation was carried out to smooth over the zeros. Another round of t-SNE revealed closer agreement between synthetic and APOGEE spectra (right image in Fig. \ref{fig:tsne}). These zero-flux values for all APOGEE spectra were fixed before predicting their stellar parameters in subsequent StarNet training from synthetic spectra. 

While interpolating the zero-point values in the observed spectra results in more accurate predictions, we caution that this may not be an ideal solution.  This is because we are modifying the data in a way that may not resemble a spectrum without the zero-point values. An alternative method would be to artificially inject zero-point values in the synthetic spectra during training.  This would allow StarNet to distribute weights more evenly across the spectrum, mimicking the training process on observed spectra. We note that this is not an issue when StarNet is trained on real APOGEE spectra that inherently have the zero-flux values.  

\subsubsection{Predictions for APOGEE DR12 Spectra}

\begin{figure}
\centering
\includegraphics[width=\linewidth]{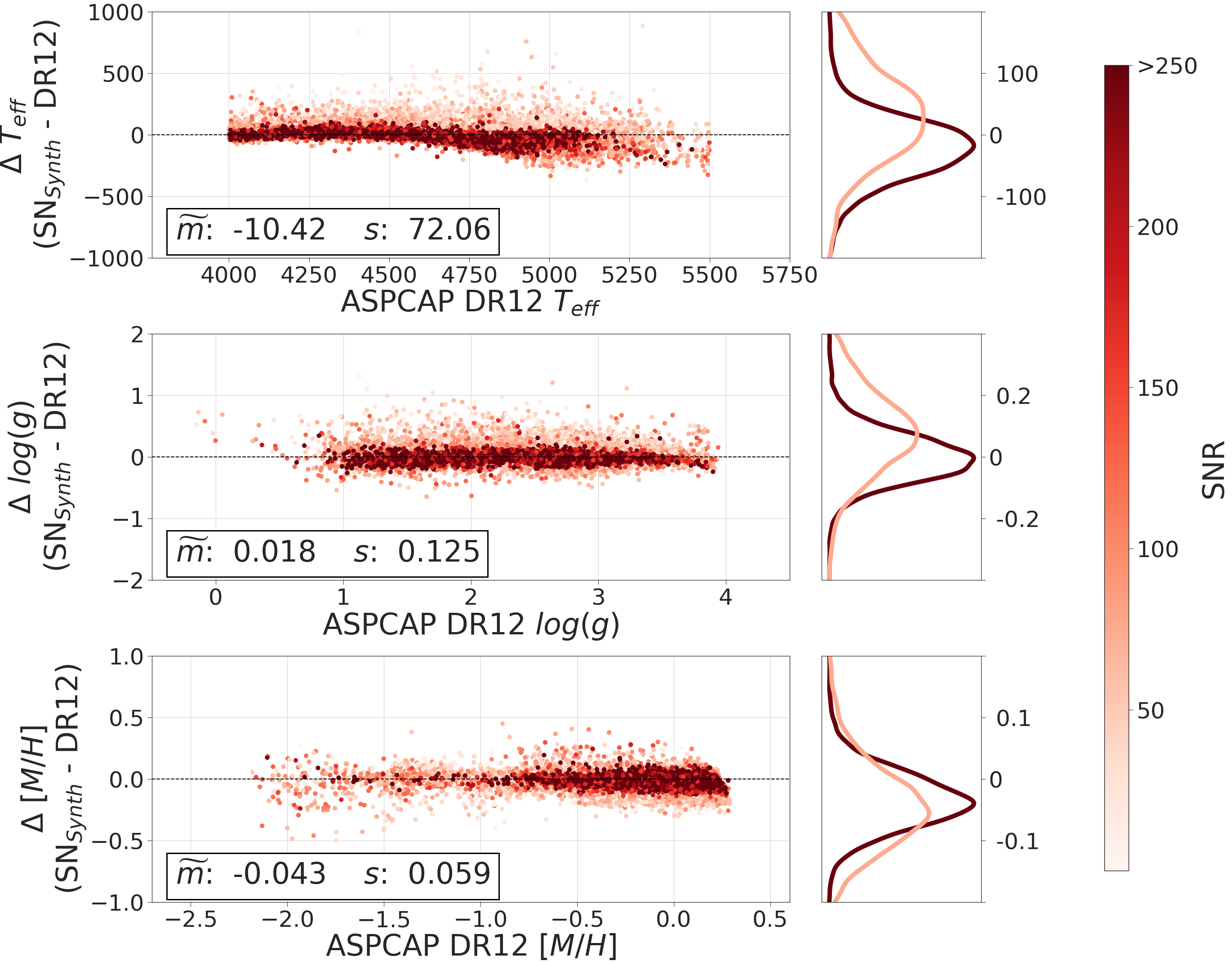}
\caption{Residuals of StarNet predictions and ASPCAP parameters for a test set of 21,787 combined spectra from APOGEE. Comparisons were made to DR12 for consistency since our tests use the ASSET synthetic grid (not used for DR13). StarNet was trained on 224,000 synthetic spectra randomly sampled from the ASSET synthetic grid by interpolating between grid points. Projected distributions about the mean are shown on the right (dark red for observational spectra with S/N $>$ 150, light red for S/N $<$ 100). The median value (\~{m}) and standard deviation (s) are calculated in each panel, as in Fig.~\ref{fig:trainsynthtestsynth}.
\label{fig:trainsynthtestreal}}
\end{figure}

After training StarNet with synthetic spectra from the ASSET code (as described in Section \ref{section:syntheticspectra}), we applied it to a dataset of 21,787 combined spectra from APOGEE DR12\footnote{The ASSET spectra are only available with continuum normalization using the ASPCAP method, therefore we adopted the ASPCAP normalized DR12 spectra to test this implementation of StarNet. The non-ASPCAP normalization scheme described in section \ref{preprocess} was only applied when StarNet was trained on APOGEE DR13 spectra} using the same restrictions found in Table \ref{table:cuts}. We compare our results to the ASPCAP DR12 parameters (see Fig. \ref{fig:trainsynthtestreal}), since those were also determined using the ASSET spectral grid (unlike the APOGEE DR13 results).
 
The distribution of the residuals is similar to those seen in Fig. \ref{HIGHSNR}, where StarNet was trained on observed APOGEE spectra. Using the method as described in Section \ref{error_propagation}, the intrinsic scatter error terms for spectra with S/N $>$ 150 were calculated to be $\Delta$\teff = 51~K,  $\Delta$\logg = 0.06,  and $\Delta$[Fe/H] = 0.08. For the entire spectral data set, including the lower S/N spectra, the intrinsic scatter errors are two times larger than when StarNet was trained on ASPCAP parameters. This is likely due to unaccounted systematic effects at training time, e.g., instrumental or extinction effects.  An extension of StarNet to include unaccounted synthetic modeling effects is planned for future work.

Interstellar extinction, atmospheric extinction, and instrumental signatures are not simulated in the synthetic spectra. These effects can vary from star to star, generally affecting the bluer regions of a spectrum as a smoothly varying function. Given that the APOGEE spectra cover infrared wavelengths for mostly nearby bright objects (H>12.5), the effects of extinction are not expected to be significant. In Section \ref{partials_section}, the sensitivity of the StarNet model to various features in the DR13 spectra is discussed.

Adding noise to the synthetic set also at the training stage allowed StarNet to learn which features impacted the stellar parameter estimates, while discounting weak features that would not be detectable in noisy APOGEE spectra. 
By varying the noise levels, it was found that a S/N $>$ 20 is necessary to decrease the residual intrinsic scatter before saturation. Adding noise to the synthetic training set not only helps to reproduce a more realistic setting, but also accounts for some of the uncertainties in the physical and instrumental modeling and decreases over-fitting. Since the APOGEE data set consists of mostly high S/N spectra, then adding a simple noise model appears to be sufficient for modeling the APOGEE spectra. However, potential applications on non-APOGEE data (or APOGEE data with more sensitive stellar parameters) may require a more thorough noise model.

\begin{figure*}
\includegraphics[width=\linewidth]{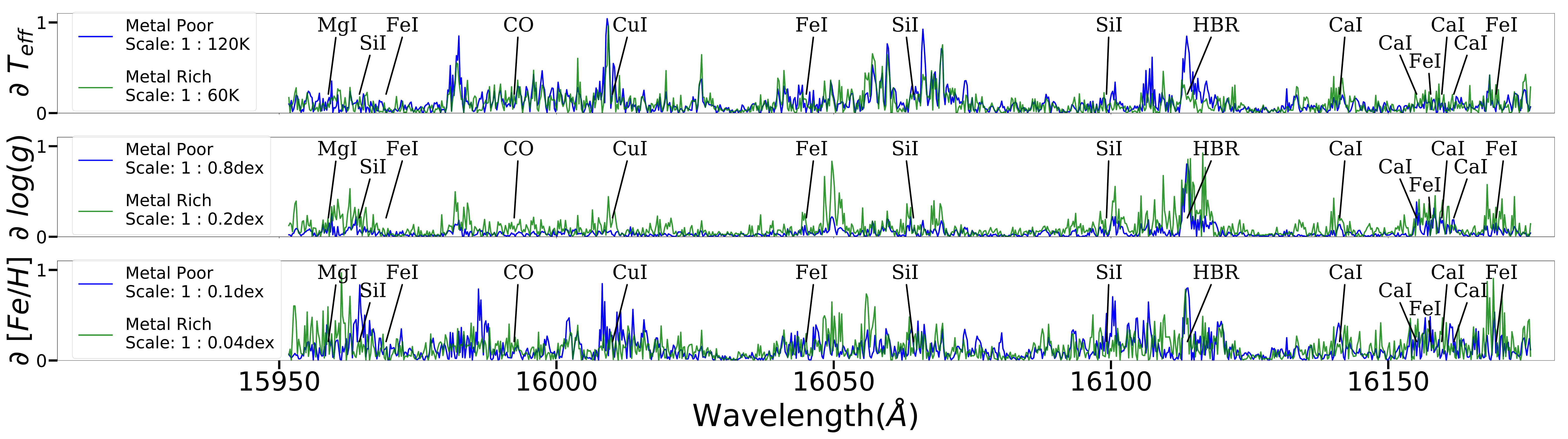}
\includegraphics[width=\linewidth]{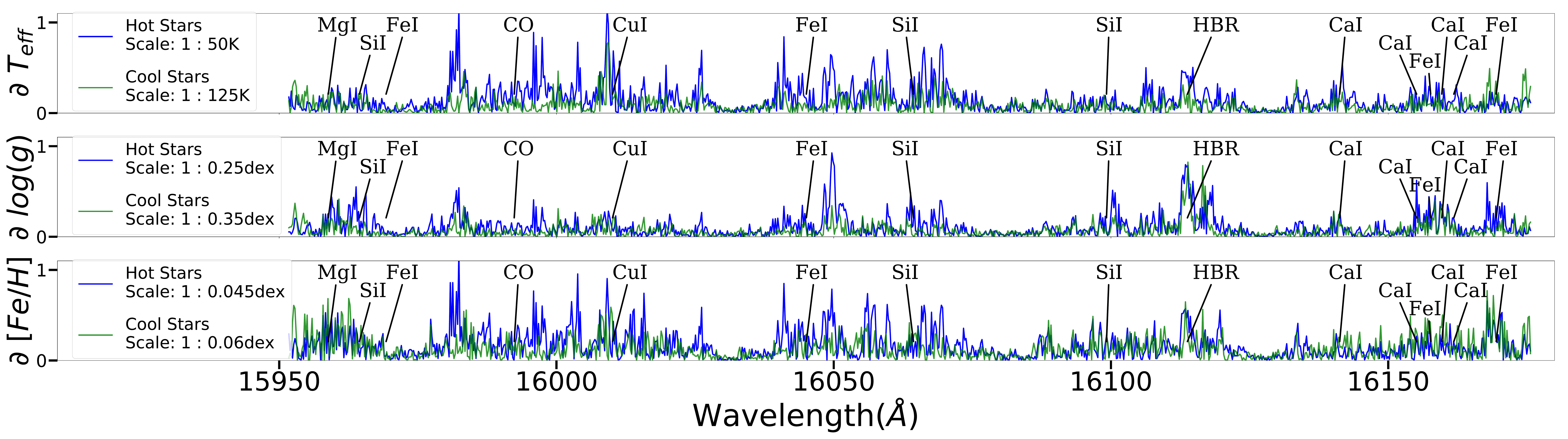}
\caption{Partial derivatives of the three stellar output parameters from the StarNet model - trained on APOGEE spectra - with respect to input wavelength bins for a section of the green chip. The partial derivatives of stars from different ranges of the parameter space were compared against each other. Stars with [Fe/H] $>$ 0.0 were compared to those with [Fe/H] $<$ -1.2 (top). Similarly stars with \teff $>$ 5000K were compare to those with \teff $<$ 4300K (bottom). An average absolute-valued Jacobian was calculated from 2000 stars in each parameter range. Note the scale differences when comparing the partial derivatives.
\label{partials_real}}
\end{figure*}

\begin{figure*}
\includegraphics[width=\linewidth]{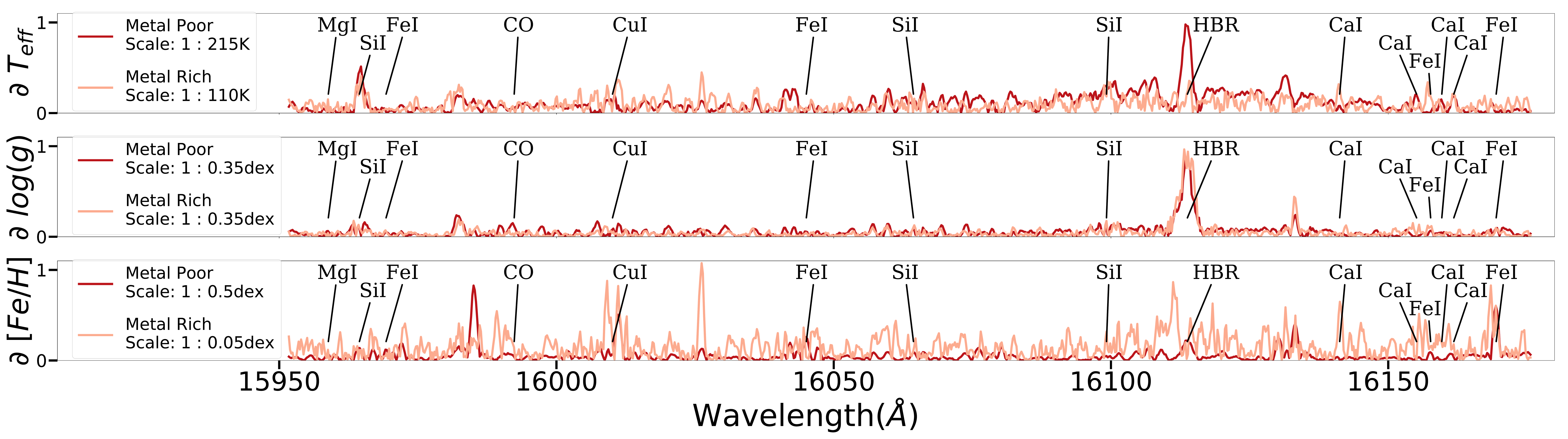}
\includegraphics[width=\linewidth]{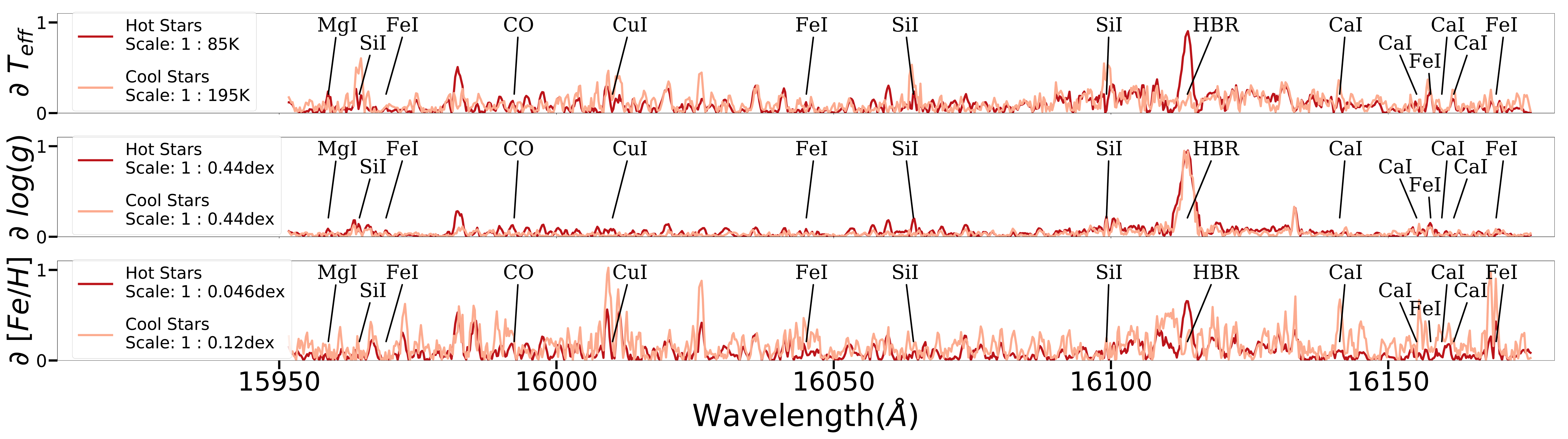}
\caption{Partial derivatives of the three stellar output parameters from the StarNet model - trained on synthetic data - with respect to input wavelength bins for a section of the green chip. The partial derivatives of stars from different ranges of the parameter space were compared against each other. Stars with [Fe/H] $>$ 0.0 were compared to those with [Fe/H] $<$ -1.2 (top). Similarly stars with \teff $>$ 5000K were compare to those with \teff $<$ 4300K (bottom). An average absolute-valued Jacobian was calculated from 2000 stars in each parameter range. Note the scale differences when comparing the partial derivatives.
\label{partials}}
\end{figure*}

\subsection{Partial Derivatives}
\label{partials_section}

It is possible to examine the learned model to determine which parts of the spectra the neural network is weighting when predicting stellar parameters. We do this by calculating the partial derivatives of each of the outputs ({\teff}, {\logg}, and [Fe/H]) with respect to every input flux value of a particular spectrum to obtain the Jacobian (as described in Section \ref{error_propagation}). 

In Fig.s \ref{partials_real} and \ref{partials} - for both StarNet trained on APOGEE DR13 spectra and synthetic spectra, respectively - we show an average of the absolute valued Jacobian from 2000 stars located in particular ranges of the parameter space. The Jacobian from metal poor stars were compared against those from metal rich stars. Comparisons were also made between hot and cool stars.
We focus on a subsection of the green chip from 15950$\mbox{\AA}$ to 16180$\mbox{\AA}$. Some features are labeled from the APOGEE input line list.
A few notable features include:

\begin{enumerate}
\item Hydrogen Bracket (HBR) lines, which play a significant role in the determination of gravity, but also in the determination of temperature in metal-poor stars
\item Atomic metal lines (e.g., FeI, CaI, CuI) that play a significant role in the determination of temperature and metallicity throughout our stellar parameter range
\item Certain atomic metal lines in the synthetic grid that play a role in the determination of temperature, e.g., SiI appears more and CaI appears less significant 
\item Unidentified or weaker lines that are not used in the APOGEE window functions and yet play an important role in the determination of all of the stellar parameters from the observed spectra
\item Unidentified or weaker lines that play a significant role only in the determination of metallicity for cool stars and metal-rich stars  
\end{enumerate}

None of these features were pre-selected or externally weighted before the neural network was trained. We also compared the partial derivatives for subsets of stars with S/N$<$60 versus S/N$>$200, but found no notable differences in the predictive power. Similar results are seen in the other wavelength regions across the whole APOGEE spectral range.

\section{Discussion}

\subsection{Optical Benchmark Stars}

The accuracy of the parameters returned by StarNet is limited by the quality of the training data. When StarNet is trained on the ASPCAP stellar parameters, uncertainties in the ASPCAP pipeline are implicitly propagated throughout the results. One option to examine the fidelity of the APOGEE StarNet results is to compare them to optical analyses. Stellar parameters and abundances for individual stars derived through optical analysis can have higher precision due to the availability of higher resolution optical spectrographs, access to more spectral features, and higher confidence in the optical stellar atmosphere models \citep[see][]{sneden2004m3m13,gilmore2012gaiaeso, venn2012, aoki2013hires, yong2013MP, howes2014gaiaesoMP, roederer2014, lamb2015chemical}.

\begin{figure*}
\centering
\includegraphics[width=\textwidth]{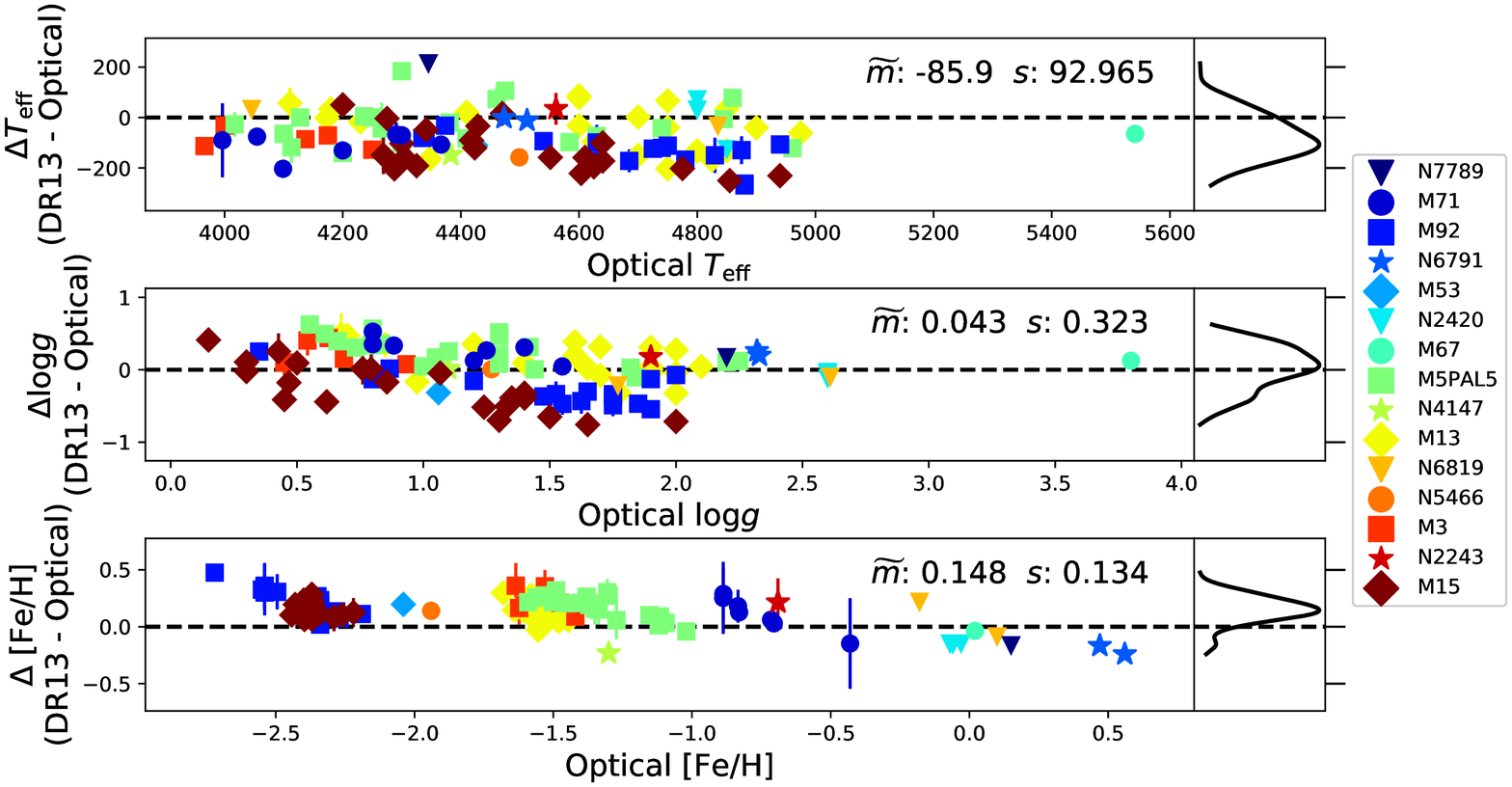}
\caption{Comparison of ASPCAP DR13 stellar parameters for 102 stars in clusters to their corresponding optical parameters sourced from the literature. When multiple literature sources are available for a particular star, the average of the reported parameters is shown with the error bars defined by the standard deviation of the optical data added in quadrature with the errors produced by StarNet. See Table \ref{tab:benchmarks} for a list of the literature references.  The median value (\~{m}) and standard deviation (s) are calculated in each panel, as in Fig.~\ref{fig:trainsynthtestsynth}.}
\label{benchmarks}
\end{figure*}

To investigate the accuracy in the StarNet predictions from training on synthetic spectra, all stars from the APOGEE DR13 database with the ``Calibration-Cluster'' or ``Standard Star'' flag were selected to create a group of benchmark stars with ASPCAP parameters.
Parameters determined from optical spectral analyses were then obtained for these stars through direct matching of the APOGEE identifier/2MASS identifier \citep{2MASS2006} to stars in the Pastel 2016 catalog \citep{Pastel2016}, and supplemented with additional data from references in \citet{meszaros2013calibrations} and \citet{holtzman2015abundances}. Fig. \ref{benchmarks} compares the ASPCAP DR13 stellar parameters to the optically determined parameters for 102 benchmark stars (tabulated in Appendix B, with references).

Examining Fig. \ref{benchmarks}, there are small offsets between the ASPCAP DR13 and the optical values for all three stellar parameters. Comparing Fig. \ref{benchmarks} to Fig. \ref{HIGHSNR} - where StarNet is trained on observed spectra and compared to ASPCAP - a similar bias at low [Fe/H] is seen in the residuals. The bias seen in Fig. \ref{HIGHSNR} is likely due to a sample size bias in the StarNet training set, though the bias in Fig. \ref{benchmarks} could be due to a insufficient coverage in the synthetic grid. These systematic biases warrant the need for larger reference samples of the parameter space, as well as possible improvements in the coverage of theoretical spectra.

Comparing the residual scatter between the DR13 ASPCAP and the StarNet predictions (Fig. \ref{HIGHSNR}) to the scatter in the optical analyses predictions (Fig. \ref{benchmarks}), we see that the optical analyses have larger scatter by a factor of $\gtrsim 2$ in \teff, $\gtrsim 4$ in \logg, and $\gtrsim 4$ in [Fe/H]. The larger scatter suggests that our simplified derivation of statistical uncertainties is sufficient until we can calibrate stellar parameters for a larger set of more homogeneous benchmark stars.  Well calibrated stellar parameters would require a more thorough probabilistic model of both StarNet training and StarNet predictions. Therefore, we consider this very good agreement, and do not investigate these intrinsic dispersions further at this time.

A comparison of Fig. \ref{fig:trainsynthtestreal} and \ref{benchmarks} reveals  that, when StarNet is trained on synthetic spectra and compared to DR12, the scatter is a factor of $\sim 2$ lower than the residuals between the DR13 and optical analyses. However, due to the different input physics, analysis techniques, spectral resolution, etc. between StarNet, APOGEE, and optical analyses, we cannot directly compare the precision of the stellar parameters returned by StarNet and an optical analysis at this time. Rather, this similar degree of scatter in the residuals indicates that the systematic errors intrinsic to StarNet are comparable to the systematic errors expected from an optical analysis. This indicates that the StarNet model is sufficient for a single survey analysis, and can provide a new and exciting tool for the efficient analyses of spectroscopic surveys.

\subsection{M Dwarfs in DR13} \label{m_dwarfs}

While APOGEE has primarily observed red giant and subgiant stars, DR13 includes spectra for known M dwarfs.  These stars are accompanied with a specific Target Flag: ``APOGEE MDWARF''.  Initially, we had not removed these stars from the dataset, and found obvious discrepancies between the StarNet results and those provided by ASPCAP DR13. On closer inspection, we found ~5900 stars that resembled the known M dwarfs, but that had not been flagged as M dwarfs.  Those stars were removed from our data sets, using one of the following cuts:

\begin{enumerate}
\item Metallicity parameters of $-0.9 <$ (ASPCAP [Fe/H]) $< 1.0$, and where (StarNet [Fe/H] $-$ ASPCAP [Fe/H]) $< -0.4$, or
\item Gravity parameters of (ASPCAP {\logg}) $> 3.6$, and where (StarNet {\logg} - ASPCAP {\logg}) $< -0.3$
\end{enumerate}

The ASSET grid provided with the APOGEE DR12 data release does not include synthetic spectra for M dwarfs, and therefore StarNet cannot be trained on those stars.  Furthermore, StarNet cannot be trained on the observed spectra at this time since there are too few M dwarfs in the  APOGEE dataset ($<$1000 flagged).  Our method for flagging M dwarfs is not robust, e.g., some RGB stars may have been removed, while other M dwarfs were not. 

\subsection{Neural Network Considerations}

\begin{figure}
\centering
\includegraphics[width=\linewidth]{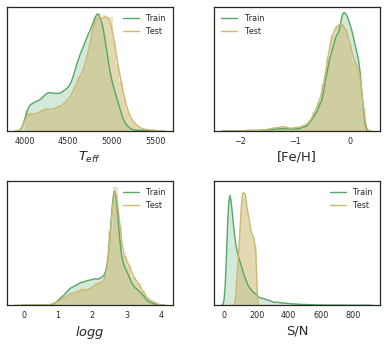}
\caption{Distribution of parameters in the StarNet training and test sets when StarNet is trained and tested on APOGEE DR13 spectra. StarNet might not test well on stars occupying the parameter space less populated by the training set.
\label{coverage}}
\end{figure}

One limitation of a deep neural network is the necessity of large training sets spanning a wide range of the parameter space. The distribution of the stellar parameters in the training set for StarNet (when trained on observed spectra) are shown in Fig. \ref{coverage}. From these plots, it is evident that there are fewer spectra for stars at high temperatures, low metallicities, and low gravities.  As seen in Figs. \ref{HIGHSNR} and \ref{cannon1_test}, having fewer training spectra in these regions of the parameter space results in less accurate predictions when testing on stars in those regions. 
StarNet is also challenged when combinations of limited parameters are scarce in the training set.  
For example, when StarNet was trained on APOGEE DR13 spectra, there were very few stars with 4500K $\leq$ {\teff} $\leq$ 5200K, 1 $\leq$ {\logg} $\leq$ 2, and [Fe/H] $\leq$ -1.6  in the reference set (see
Fig. \ref{veracity}). This issue could be mitigated by adjusting the restrictions set on the StarNet reference set, but may cost in accuracy.

While stellar observations are limited by time and resources, generating synthetic spectra are limited only by computing resources. Being able to produce synthetic spectra that evenly span a parameter space can improve the ability of a neural network to predict in regions which may be affected by a sample bias. We see this as one of the strengths of StarNet.  

In the future, we envision combining observed spectra with synthetically generated spectra to improve the precision in the StarNet predictions.   This would increase the number of training spectra over the entire parameter range, while also reducing the effect of the synthetic gap.   Of course, this NN application is not limited to the APOGEE wavelength region, but can be extended to other spectroscopic surveys.

Finally, some skepticism exists around machine learning due to its historical inaccessibility and complexity; sometimes being perceived as a black box. New developments have improved usability and made machine learning much more accessible. The training and testing of all of the models used in this paper were performed with the neural network library, \texttt{KERAS} \citep{chollet2015keras}, which provides a high level application program interface to the \texttt{TENSORFLOW} \citep{tensorflow2015-whitepaper} machine intelligence software package. We provide code to reproduce steps of training and testing on the APOGEE DR13 data set with StarNet at https://github.com/astroai/starnet.

\section{Conclusion}

We have presented a convolutional neural network model that is capable of determining the stellar parameters \teff, \logg, and [Fe/H] directly from spectra.   Our development of this model, StarNet, includes training and testing on real spectra (from the APOGEE database) and synthetic spectra (from the APOGEE ASSET grid).  We have implemented an optimization procedure for hyper-parameter tuning, examining the precision in the stellar parameter predictions when developing the StarNet architecture.

By applying this StarNet model to various data releases from the APOGEE survey, we show that it is capable of estimating stellar parameters from {\it synthetic spectra}.  These parameters and the uncertainties are similar to the ASPCAP results.  We also compare StarNet to other {\it data-driven} methods (such as The Cannon 2) and find similar results when trained on sufficiently large observational data sets. 

In our applications on the APOGEE data sets, we also explore the limitations of the training sets; the sample sizes; the stellar parameter ranges; the impact of zero-flux values in real spectra during training and during testing; and the effect of signal-to-noise of the spectra on the accuracy of the StarNet predictions. Our results are robust over a large range of S/N values for the APOGEE spectra, including low S/N ($\sim$15).

In the future, testing of the intrinsic errors in neural network models should be further examined, as well as variations in the synthetic spectral grids; e.g., variations in the model atmospheres adopted when generating the synthetic spectra, changes in the continuum normalization methods, as well as the exploration of different wavelength regions.   We are especially interested in applying StarNet to optical spectroscopic surveys and observations.

\section*{Acknowledgements}
We thank the anonymous referee for helpful comments that have improved this manuscript. 
This paper was first presented at l'Observatoire de la C{\^o}te d'Azur in Nice, France on 1 June 2017, and again (in part) at the Joint Institute for Nuclear Astrophysics Forging Connections meeting in Lansing, Michigan on 28 June 2017.
The authors wish to thank Dr. Mike Irwin (Institute of Astronomy, Univ. of Cambridge) and Dr. Jo Bovy (University of Toronto) for helpful comments on this work, and Drs. Vanessa Hill, Alejandra Recio-Blanco, and Patrick de Laverny (Observatoire de la C{\^o}te d'Azur, France) for many interesting discussions on spectral analyses.
SB and TO thank the National Research Council Herzberg Astronomy and Astrophysics for funding through their co-op program. KAV, CLK, FJ, SB, TO, and SM acknowledge partial funding for this work through a National Sciences and Engineering Research Council Collaborative Research and Training Experience Program in {\it New Technologies for Canadian Observatories}.




\bibliographystyle{mnras}
\bibliography{Starnet} 


\onecolumn
\appendix

\section{The StarNet Convolutional Neural Network}
\label{appendix_method}

In this paper, we approach the prediction of stellar parameters from spectra with supervised learning. From a data set consisting of individual stellar spectra and associated physical parameters, we approximate a function to learn the mapping between the two. For other similar spectra, we then assume the trained model is capable of predicting stellar parameters. 

Our mapping function is parameterized and organized as a neural network (NN). It consists of a collection of artificial neurons or nodes arranged in layers: an \textit{input layer} (in our case the input spectra), a number of \textit{hidden layers}, and an \textit{output layer} (in our case the stellar parameters). Each node is parameterized as a linear function of weights, \qvec{w}, applied to an input, \qvec{x}, with an additional bias value, b. A node is then activated by a nonlinear function, $g$, giving an output of a node to be written as:
\begin{equation*}
h(\qvec{x}) = g(\qvec{w$^T$x} + b)
\end{equation*}

\noindent Common activation functions include the sigmoid function and the Rectified Linear Unit (ReLU):
\begin{equation*}
g(z) = \max(0, z)
\end{equation*}

\noindent
which allow the network to adapt to non-linear problems \citep{chen1990non}. The ReLU function was used in the implementation for all StarNet layers except for the last output layer.

In a traditional sequential NN architecture, each node is connected to every node from the previous layer as well as every node in the following layer, thereafter referred to as \textit{fully connected} layers. At hidden layer, $l$, the output, \qvec{h}$^{(l)}$, is a vector valued function of the previous layer, \qvec{h}$^{(l-1)}$, and is given by:
\begin{equation*}
\qvec{h}^{(l)} = \qvec{g}(\qvec{w}^{(l)} \qvec{h}^{(l-1)} + \qvec{b}^{(l)})
\end{equation*}

\noindent The first layer is simply the input \qvec{h}$^{(0)}$(\qvec{x}) = \qvec{x}, in our case: the spectra.  The next two layers of StarNet are \textit{convolutional layers}, which are more adapted to higher dimensional inputs by leveraging local connectivity in the previous layer. In convolutional layers, the weights are applied as filters. The filter slides across the previous layer taking the dot product between the filter weights and sections of the input. For a given filter covering a section, $s$, this operation can be summarized as:
\begin{equation*}
\qvec{h}^{(l)}_s = \qvec{g}(\qvec{w$_s$}^{(l)} \otimes \qvec{h}^{(l-1)}+ \qvec{b$_s$})
\end{equation*}

\noindent These filters allow for the extraction of features in the input and learn which features to extract through training. After the convolutional layers in StarNet, we use a max pooling layer. A max pooling layer is a non-linear down-sampling technique typically used in convolutional neural networks to decrease the number of free parameters and to extract the strongest features from a convolutional layer. In a max pooling layer, a window moves along the feature map generated by each of the filters in the previous convolutional layer - in strides of length equal to the length of the window - extracting the maximum value from each sub-region. These pools of maxima are then passed on to the following layer. The next two layers in StarNet are fully-connected layers.

The combination of all those layers allows for the formation of non-linear combinations of an input vector, \qvec{x}, to produce an output vector prediction, ${f}$(\qvec{x; w}, $b$). For each training sample of spectra, \qvec{x$_t$}, and corresponding known stellar parameters, \qvec{y$_t$}, the NN model weights and biases are estimated by minimizing the \textit{empirical risk} that computes the \textit{loss} between the predictions and targets for a batch of $T$ training samples, often supplemented with a regularizing function. We ended up adopting a mean-squared-error loss function without regularization for StarNet, such that the StarNet empirical risk to be minimized reads:
\begin{equation*}
\argmin_{\qvec{w},b} \frac{1}{T}\sum_{t=1}^T \big(\qvec{y$_t$} - f(\qvec{x$_t$; w}, b)\big)^2.
\end{equation*}

The minimization is performed with a stochastic gradient descent (SGD) algorithm. SGD algorithms require the computation of the loss function gradients with respect to the weights, and make adjustments to those weights iteratively until reaching a minimum. In our case, the optimization is performed using the ADAM optimizer \citep{kingma2014adam}, an SGD variant using adaptive estimates of the gradient moments to adjust learning rates. Initially, the weights of the model are randomly set and therefore the predictions will be quite poor. Computing the gradients is operated backwards through each sequential layer, a process referred as \textit{back-propagation}, and is the computationally expensive part of the training. 

In the case of StarNet, a cross-validation set was used to test the model following every iteration to evaluate whether or not the model had improved; if improvements were not made after a given number of iterations, the training was stopped. This minimum may differ depending on the complexity of the model architecture as well as various hyper-parameters (discussed in Section \ref{model_comparison}). Following each iteration, the cross-validation set is sent through a single forward propagation where the outputs are predicted and compared against the target values. This set is not used for training, but only to ensure that the model is not over-fitting to the training set. Over-fitting occurs when a model learns a function that may be able to compute the outputs for the training set very well, but can not generalize that function to be applied to a test set that is not included in the training. A cross-validation set is used as a type of middle-ground between the training and test set, and ensures that over-fitting does not occur. If the cross-validation predictions do not improve after several iterations, the training will be stopped. Using a cross-validation set to avoid over-fitting and tuning hyper-parameters is common practice in machine learning applications \citep{gurney1997introduction}.

Once a minimum has been reached, a separate test set - not included in training or cross-validation - is selected to ensure that the model performs well on new data. For StarNet, this test set consists of stellar spectra with ``known'' stellar parameters that are used to compare with StarNet predictions.
For more in depth explanation of deep neural networks, we refer the reader to \citet{deeplearningbook}.

\section{Optical benchmark stars in clusters}
In this Appendix, we tabulate the stellar parameters for stars in globular clusters that have both APOGEE DR13 results (based on IR spectral analyses) and results in the literature (based on high-resolution optical spectral analyses).  Our systematic investigation has revealed numerous stars with incomplete optical parameters in the literature (e.g., missing $T_{\rm eff}$, $\log g$, or [Fe/H]), or incomplete IR parameters in the ASPCAP database (e.g., presumably due to data acquisition issues such as low $S/N$ or persistence).  We have removed those stars from our benchmark sample, as well as targets in the APOGEE database with a high RV scatter between individual visits (which usually indicates a potential binary system).  
Fig. \ref{benchmarks} compares the stellar parameters for our final sample of optical benchmark stars (as listed below, with the obvious outliers removed). Small systematic offsets in $T_{\rm eff}$, $\log g$, and [Fe/H] are found, though typically within the quoted 1$\sigma$ errors. 

\begin{landscape}
\begin{scriptsize}  

\begin{longtable}{@{\extracolsep{\fill}}cccccccccccccccc@{}}
\label{tab:benchmarks}

 2MASS ID  & Cluster &
 \begin{tabular}[c]{@{}c@{}}APOGEE\\ PERS\\ Flag\end{tabular} &
  \multicolumn{3}{c}{ASPCAP DR13 Parameters} &
   \multicolumn{3}{c}{Optical Parameters} &
    \multicolumn{3}{c}{StarNet$_{\mathrm{ASPCAP}}$} &
     \multicolumn{3}{c}{StarNet$_{\mathrm{Synth}}$} &
     References \\
     &  &   & T$_{\mathrm{eff}}$ & $\log g$  & [Fe/H] & 
     T$_{\mathrm{eff}}$  & $\log g$ & [Fe/H] & 
     T$_{\mathrm{eff}}$  & $\log g$ & [Fe/H] &
     T$_{\mathrm{eff}}$  & $\log g$ & [Fe/H]  & \\
     &  &  & [K]  & [dex]  & [dex] & [K] & [dex] & [dex] & [K] & [dex] & [dex] & [K] & [dex] & [dex] \\ \hline

\multicolumn{2}{c}{Representative Errors}       &                                & 92         & 0.11      & 0.05     & 86                         & 0.24                      & 0.17                     & 47           & 0.14       & 0.07       &    &  & &                          \\
2M16412709+3628002  &  M13  &  --  &  4184  &  0.81  &  -1.54  &  4348  &  0.98  &  -1.54  &  4201  &  0.96  &  -1.48  &  4405  &  0.99  &  -1.61  &  h, ac, af, al \\
2M16413053+3629434  &  M13  &  --  &  4431  &  1.55  &  -1.39  &  4410  &  1.2  &  -1.44  &  4434  &  1.49  &  -1.34  &  4429  &  1.47  &  -1.42  &  ac, af \\
2M16413072+3630075  &  M13  &  High  &  4816  &  2.21  &  -1.29  &  4750  &  1.9  &  -1.5  &  4809  &  2.2  &  -1.27  &  4742  &  1.9  &  -1.36  &  k \\
2M16413082+3630130  &  M13  &  --  &  4658  &  1.45  &  -1.58  &  4825  &  1.77  &  -1.55  &  4665  &  1.74  &  -1.56  &  4869  &  1.98  &  -1.6  &  k, ag \\
2M16413476+3627596  &  M13  &  --  &  4213  &  1.2  &  -1.33  &  4230  &  0.85  &  -1.44  &  4205  &  1.14  &  -1.3  &  4249  &  1.02  &  -1.34  &  ad, af \\
2M16413482+3627197  &  M13  &  High  &  4171  &  1.17  &  -1.34  &  4173  &  0.7  &  -1.49  &  4157  &  1.12  &  -1.3  &  4220  &  1.0  &  -1.37  &  j, ad, af \\
2M16413684+3629289  &  M13  &  --  &  4547  &  1.68  &  -1.43  &  4750  &  2.0  &  -1.48  &  4582  &  1.89  &  -1.5  &  4869  &  2.25  &  -1.53  &  ag \\
2M16413707+3630378  &  M13  &  --  &  4550  &  1.62  &  -1.48  &  4700  &  1.7  &  -1.58  &  4597  &  1.87  &  -1.49  &  4829  &  2.16  &  -1.51  &  ag \\
2M16413870+3625380  &  M13  &  High  &  4168  &  1.17  &  -1.31  &  4111  &  0.67  &  -1.56  &  4134  &  1.09  &  -1.29  &  4205  &  0.95  &  -1.33  &  f, j, ac, ad, af, al \\
2M16413945+3632029  &  M13  &  --  &  4883  &  2.14  &  -1.41  &  4850  &  2.1  &  -1.56  &  4839  &  2.21  &  -1.44  &  4970  &  2.47  &  -1.45  &  ag \\
2M16414398+3622338  &  M13  &  High  &  4568  &  1.49  &  -1.47  &  4600  &  1.4  &  -1.55  &  4638  &  1.62  &  -1.45  &  4712  &  1.63  &  -1.47  &  k \\
2M16414478+3623273  &  M13  &  High  &  4529  &  1.67  &  -1.47  &  4625  &  1.65  &  -1.6  &  4522  &  1.57  &  -1.49  &  4657  &  1.58  &  -1.59  &  k, ag \\
2M16414528+3631068  &  M13  &  --  &  4679  &  1.8  &  -1.41  &  4600  &  1.6  &  -1.57  &  4742  &  2.0  &  -1.34  &  4701  &  2.07  &  -1.35  &  k, ag \\
2M16414558+3630328  &  M13  &  --  &  4711  &  1.86  &  -1.43  &  4750  &  1.9  &  -1.57  &  4781  &  2.19  &  -1.35  &  4733  &  2.18  &  -1.37  &  k \\
2M16414744+3628284  &  M13  &  --  &  4214  &  1.21  &  -1.34  &  4180  &  0.8  &  -1.46  &  4203  &  1.14  &  -1.32  &  4261  &  1.14  &  -1.34  &  ad, af \\
2M16414932+3625264  &  M13  &  High  &  4686  &  1.99  &  -1.3  &  4600  &  1.6  &  -1.58  &  4736  &  2.05  &  -1.3  &  4741  &  1.92  &  -1.27  &  k \\
2M16415037+3623417  &  M13  &  High  &  4702  &  1.9  &  -1.4  &  4700  &  1.9  &  -1.54  &  4771  &  1.94  &  -1.35  &  4697  &  1.76  &  -1.37  &  ag \\
2M16415160+3629363  &  M13  &  High  &  4914  &  2.02  &  -1.38  &  4975  &  1.7  &  -1.68  &  4782  &  1.95  &  -1.51  &  4923  &  1.62  &  -1.61  &  ag \\
2M16415239+3628395  &  M13  &  High  &  4668  &  1.96  &  -1.39  &  4800  &  2.0  &  -1.52  &  4631  &  1.85  &  -1.52  &  4858  &  2.04  &  -1.46  &  k \\
2M16415842+3628312  &  M13  &  High  &  4860  &  2.28  &  -1.32  &  4900  &  2.0  &  -1.53  &  4754  &  2.18  &  -1.36  &  4901  &  2.05  &  -1.48  &  k \\
2M16415862+3627465  &  M13  &  High  &  4719  &  1.94  &  -1.49  &  4850  &  1.9  &  -1.64  &  4618  &  1.89  &  -1.43  &  5002  &  2.35  &  -1.52  &  ag \\
2M13131736+1814463  &  M53  &  --  &  4314  &  0.74  &  -1.84  &  4425  &  1.06  &  -2.04  &  4414  &  1.04  &  -1.79  &  4620  &  1.19  &  -1.9  &  ah \\
2M07380627+2136542  &  N2420  &  --  &  4725  &  2.55  &  -0.22  &  4850  &  2.6  &  -0.07  &  4725  &  2.56  &  -0.22  &  4696  &  2.57  &  -0.22  &  am \\
2M07381549+2138015  &  N2420  &  --  &  4872  &  2.57  &  -0.21  &  4800  &  2.6  &  -0.06  &  4890  &  2.57  &  -0.21  &  4855  &  2.71  &  -0.18  &  am \\
2M07382696+2138244  &  N2420  &  --  &  4832  &  2.48  &  -0.18  &  4800  &  2.6  &  -0.03  &  4825  &  2.48  &  -0.17  &  4824  &  2.65  &  -0.16  &  am \\
2M12100405+1832532  &  N4147  &  High  &  4235  &  1.12  &  -1.53  &  4383  &  1.1  &  -1.3  &  4270  &  0.92  &  -1.73  &  4458  &  0.79  &  -1.86  &  ak \\
2M13414576+2824597  &  M3  &  --  &  4051  &  0.84  &  -1.46  &  4137  &  0.69  &  -1.62  &  4057  &  0.89  &  -1.46  &  4277  &  0.76  &  -1.52  &  j, r, x, y, ac, af, al \\
2M13421204+2826265  &  M3  &  --  &  3968  &  0.95  &  -1.27  &  4000  &  0.54  &  -1.64  &  3975  &  0.87  &  -1.27  &  4172  &  0.55  &  -1.42  &  r, x, y, ac, af, al \\
2M13421679+2823479  &  M3  &  --  &  4123  &  1.01  &  -1.33  &  4250  &  0.93  &  -1.42  &  4119  &  0.95  &  -1.32  &  4260  &  0.97  &  -1.4  &  j, ac, af \\
2M13423922+2827574  &  M3  &  Low  &  4104  &  1.06  &  -1.27  &  4175  &  0.63  &  -1.48  &  4110  &  1.03  &  -1.3  &  4263  &  0.94  &  -1.35  &  x, z \\
2M13424150+2819081  &  M3  &  --  &  3853  &  0.55  &  -1.17  &  3966  &  0.45  &  -1.53  &  3910  &  0.98  &  -1.12  &  4110  &  0.51  &  -1.27  &  r, x, y, ac, af \\
2M14052071+2829419  &  N5466  &  High  &  4341  &  1.28  &  -1.8  &  4499  &  1.27  &  -1.94  &  4472  &  1.38  &  -1.67  &  4846  &  1.76  &  -1.87  &  ah \\
2M08514401+1146245  &  M67  &  High  &  5476  &  3.93  &  -0.02  &  5541  &  3.8  &  0.02  &  5363  &  3.62  &  -0.1  &  5242  &  3.7  &  -0.13  &  i \\
2M15174702+0204519  &  M5PAL5  &  --  &  4361  &  1.22  &  -1.25  &  4381  &  1.05  &  -1.53  &  4362  &  1.27  &  -1.25  &  4528  &  1.47  &  -1.28  &  u \\
2M15180831+0158530  &  M5PAL5  &  --  &  4840  &  1.73  &  -1.37  &  4961  &  1.84  &  -1.59  &  4861  &  1.9  &  -1.39  &  5164  &  2.36  &  -1.25  &  u \\
2M15180987+0210088  &  M5PAL5  &  Low  &  4557  &  1.45  &  -1.3  &  4630  &  1.44  &  -1.56  &  4558  &  1.53  &  -1.32  &  4751  &  1.77  &  -1.32  &  u \\
2M15181075+0212356  &  M5PAL5  &  Low  &  4698  &  1.85  &  -1.18  &  4740  &  1.82  &  -1.37  &  4710  &  1.99  &  -1.13  &  4843  &  2.43  &  -1.13  &  u \\
2M15181867+0204327  &  M5PAL5  &  Low  &  3995  &  1.04  &  -1.05  &  4113  &  0.73  &  -1.15  &  3996  &  1.09  &  -1.06  &  4184  &  0.69  &  -1.19  &  l, af, ak \\
2M15182262+0200305  &  M5PAL5  &  --  &  4035  &  1.05  &  -1.03  &  4100  &  0.66  &  -1.3  &  4034  &  1.11  &  -1.02  &  4203  &  0.94  &  -1.12  &  u \\
2M15182283+0203097  &  M5PAL5  &  --  &  4251  &  1.35  &  -1.11  &  4250  &  1.1  &  -1.12  &  4234  &  1.3  &  -1.11  &  4352  &  1.43  &  -1.09  &  l, af \\
2M15182345+0159572  &  M5PAL5  &  --  &  4483  &  1.76  &  -1.02  &  4300  &  1.3  &  -1.11  &  4459  &  1.69  &  -1.01  &  4381  &  1.61  &  -1.05  &  l, af \\
2M15182591+0205076  &  M5PAL5  &  --  &  4198  &  1.37  &  -1.06  &  4300  &  0.8  &  -1.02  &  4175  &  1.23  &  -1.07  &  4283  &  1.31  &  -1.09  &  l, af \\
2M15183223+0201341  &  M5PAL5  &  --  &  4060  &  1.05  &  -1.06  &  4200  &  1.0  &  -1.09  &  4053  &  1.13  &  -1.06  &  4236  &  0.98  &  -1.13  &  l, af \\
2M15183531+0207400  &  M5PAL5  &  --  &  4327  &  1.51  &  -1.19  &  4409  &  1.3  &  -1.35  &  4340  &  1.45  &  -1.19  &  4487  &  1.7  &  -1.21  &  af, aj, ak \\
2M15183738+0206079  &  M5PAL5  &  Low  &  4241  &  1.1  &  -1.17  &  4236  &  0.61  &  -1.49  &  4240  &  1.08  &  -1.2  &  4332  &  1.09  &  -1.19  &  u \\
2M15183765+0201212  &  M5PAL5  &  --  &  4839  &  2.31  &  -1.25  &  4845  &  2.2  &  -1.46  &  4835  &  2.28  &  -1.27  &  5132  &  2.7  &  -1.22  &  u \\
2M15184132+0205014  &  M5PAL5  &  --  &  4534  &  1.73  &  -1.19  &  4460  &  1.42  &  -1.34  &  4534  &  1.82  &  -1.2  &  4589  &  1.96  &  -1.16  &  u, af, aj \\
2M15184164+0203533  &  M5PAL5  &  --  &  4128  &  1.19  &  -1.07  &  4128  &  0.83  &  -1.11  &  4103  &  1.09  &  -1.07  &  4216  &  1.05  &  -1.13  &  l, af, ak \\
2M15184346+0203074  &  M5PAL5  &  --  &  4581  &  1.82  &  -1.12  &  4475  &  1.3  &  -1.38  &  4615  &  1.91  &  -1.13  &  4540  &  1.9  &  -1.1  &  u \\
2M15184374+0208171  &  M5PAL5  &  Low  &  4938  &  2.37  &  -1.19  &  4860  &  2.25  &  -1.4  &  4865  &  2.29  &  -1.18  &  5036  &  3.15  &  -1.04  &  u \\
2M15184495+0202034  &  M5PAL5  &  --  &  3990  &  1.17  &  -1.0  &  4017  &  0.55  &  -1.3  &  3981  &  1.01  &  -1.02  &  4123  &  0.79  &  -1.14  &  ak, al \\
2M15184540+0204302  &  M5PAL5  &  --  &  4223  &  1.14  &  -1.21  &  4266  &  0.8  &  -1.27  &  4222  &  1.11  &  -1.25  &  4363  &  1.09  &  -1.25  &  l, u, af, ak, al \\
2M15185515+0214337  &  M5PAL5  &  --  &  4487  &  1.38  &  -1.28  &  4584  &  1.3  &  -1.5  &  4489  &  1.51  &  -1.31  &  4669  &  1.75  &  -1.29  &  u \\
2M17163427+4307363  &  M92  &  --  &  4638  &  1.19  &  -2.24  &  4750  &  1.63  &  -2.54  &  4791  &  1.78  &  -1.86  &  4771  &  1.41  &  -2.14  &  ab, ae \\
2M17165185+4308031  &  M92  &  --  &  4620  &  1.41  &  -2.24  &  4740  &  1.75  &  -2.72  &  4761  &  1.73  &  -1.81  &  4765  &  1.98  &  -2.03  &  ae \\
2M17165557+4309277  &  M92  &  --  &  4264  &  0.6  &  -2.19  &  4340  &  0.35  &  -2.27  &  4382  &  1.05  &  -1.93  &  4454  &  0.74  &  -2.3  &  x \\
2M17165738+4307236  &  M92  &  --  &  4254  &  0.67  &  -2.15  &  4335  &  0.8  &  -2.28  &  4352  &  1.03  &  -1.95  &  4430  &  0.81  &  -2.28  &  n, x, ab, af \\
2M17165772+4314115  &  M92  &  --  &  4443  &  1.11  &  -2.21  &  4615  &  1.48  &  -2.34  &  4515  &  1.34  &  -1.89  &  4612  &  1.41  &  -2.24  &  ab, ae \\
2M17165883+4315116  &  M92  &  --  &  4513  &  1.08  &  -2.2  &  4685  &  1.55  &  -2.34  &  4746  &  1.65  &  -1.78  &  4654  &  1.27  &  -1.99  &  ab, ae \\
2M17165956+4306456  &  M92  &  --  &  4618  &  1.35  &  -2.32  &  4880  &  1.9  &  -2.34  &  4723  &  1.63  &  -1.88  &  4979  &  1.99  &  -2.31  &  ab \\
2M17165967+4301058  &  M92  &  High  &  4746  &  1.77  &  -2.11  &  4875  &  1.9  &  -2.34  &  4803  &  1.87  &  -1.73  &  4904  &  2.32  &  -1.92  &  ab, ae \\
2M17170033+4311478  &  M92  &  --  &  4680  &  1.26  &  -2.19  &  4830  &  1.75  &  -2.54  &  4677  &  1.54  &  -1.86  &  4817  &  1.77  &  -2.0  &  ab, ae \\
2M17170043+4305117  &  M92  &  --  &  4601  &  1.35  &  -2.19  &  4725  &  1.65  &  -2.5  &  4727  &  1.77  &  -1.84  &  4626  &  1.56  &  -2.08  &  ab, ae \\
2M17170081+4310251  &  M92  &  --  &  4531  &  1.19  &  -2.18  &  4630  &  1.52  &  -2.54  &  4655  &  1.6  &  -1.88  &  4679  &  1.52  &  -2.08  &  ab, ae \\
2M17170647+4306029  &  M92  &  --  &  4613  &  1.37  &  -2.22  &  4780  &  1.75  &  -2.55  &  4679  &  1.61  &  -1.88  &  4812  &  1.88  &  -2.06  &  ae \\
2M17170731+4309308  &  M92  &  --  &  4447  &  1.05  &  -2.16  &  4540  &  1.2  &  -2.34  &  4532  &  1.33  &  -1.9  &  4542  &  1.16  &  -2.11  &  ab \\
2M17171221+4302209  &  M92  &  High  &  4833  &  1.93  &  -2.08  &  4940  &  2.0  &  -2.35  &  4768  &  1.83  &  -1.73  &  4920  &  2.28  &  -1.9  &  ae \\
2M17171307+4309483  &  M92  &  --  &  4341  &  0.88  &  -2.08  &  4373  &  0.87  &  -2.19  &  4470  &  1.28  &  -1.88  &  4385  &  0.91  &  -2.0  &  n, ab, af \\
2M17172166+4311031  &  M92  &  --  &  4611  &  1.38  &  -2.2  &  4880  &  1.85  &  -2.39  &  4693  &  1.63  &  -1.84  &  4888  &  2.03  &  -2.03  &  ae \\
2M21290843+1209118  &  M15  &  --  &  4467  &  1.08  &  -2.14  &  4640  &  1.4  &  -2.37  &  4595  &  1.35  &  -1.91  &  4507  &  1.09  &  -2.0  &  ab \\
2M21294465+1207307  &  M15  &  --  &  4451  &  0.96  &  -2.19  &  4610  &  1.35  &  -2.37  &  4570  &  1.26  &  -1.88  &  4600  &  1.34  &  -2.08  &  ab \\
2M21294979+1211058  &  M15  &  Low  &  4303  &  0.69  &  -2.16  &  4423  &  0.86  &  -2.26  &  4475  &  1.15  &  -1.94  &  4420  &  0.75  &  -2.12  &  e, n, n, o, o, q, ab, af, af \\
2M21295311+1212310  &  M15  &  --  &  4382  &  0.83  &  -2.22  &  4604  &  1.33  &  -2.34  &  4468  &  1.25  &  -1.98  &  4596  &  1.11  &  -2.21  &  d, ab \\
2M21295492+1213225  &  M15  &  Low  &  4145  &  0.29  &  -2.27  &  4305  &  0.47  &  -2.39  &  4303  &  0.82  &  -1.95  &  4405  &  0.66  &  -2.36  &  d, m, o, o, ab \\
2M21295560+1212422  &  M15  &  --  &  4395  &  0.8  &  -2.18  &  4429  &  0.79  &  -2.38  &  4498  &  1.13  &  -1.9  &  4542  &  1.12  &  -2.03  &  d, o, o \\
2M21295562+1210455  &  M15  &  --  &  4250  &  0.56  &  -2.16  &  4200  &  0.15  &  -2.37  &  4387  &  1.04  &  -1.92  &  4376  &  0.63  &  -2.1  &  o, o, ab \\
2M21295618+1212337  &  M15  &  --  &  4327  &  0.77  &  -2.21  &  4418  &  0.76  &  -2.29  &  4489  &  1.17  &  -1.92  &  4456  &  0.9  &  -2.09  &  n, n, o, o, af \\
2M21295666+1209463  &  M15  &  High  &  4120  &  0.28  &  -2.34  &  4269  &  0.3  &  -2.44  &  4274  &  0.81  &  -1.94  &  4381  &  0.66  &  -2.33  &  m, o, o, ab \\
2M21295678+1210269  &  M15  &  --  &  4134  &  0.04  &  -2.27  &  4325  &  0.45  &  -2.42  &  4310  &  0.76  &  -1.98  &  4349  &  0.33  &  -2.22  &  o, o \\
2M21295801+1214260  &  M15  &  --  &  4605  &  1.29  &  -2.26  &  4855  &  2.0  &  -2.37  &  4569  &  1.32  &  -1.95  &  4929  &  2.03  &  -2.19  &  ab \\
2M21295856+1209214  &  M15  &  --  &  4197  &  0.41  &  -2.23  &  4300  &  0.3  &  -2.43  &  4373  &  0.97  &  -1.97  &  4392  &  0.73  &  -2.14  &  m, o, o, ab \\
2M21300033+1210508  &  M15  &  --  &  4292  &  0.68  &  -2.17  &  4341  &  0.43  &  -2.4  &  4427  &  1.08  &  -1.94  &  4434  &  0.83  &  -2.12  &  d, o, o \\
2M21300038+1207363  &  M15  &  --  &  4394  &  0.72  &  -2.26  &  4552  &  1.24  &  -2.4  &  4546  &  1.16  &  -2.0  &  4628  &  1.05  &  -2.22  &  d, ab \\
2M21300224+1211215  &  M15  &  Low  &  4084  &  0.18  &  -2.29  &  4288  &  0.62  &  -2.39  &  4255  &  0.78  &  -1.96  &  4371  &  0.59  &  -2.32  &  d, o, o, ab \\
2M21300274+1210438  &  M15  &  High  &  4271  &  0.59  &  -2.24  &  4275  &  0.5  &  -2.39  &  4432  &  1.05  &  -1.94  &  4389  &  0.66  &  -2.16  &  o, o, ab \\
2M21300637+1206592  &  M15  &  High  &  4430  &  0.6  &  -2.33  &  4625  &  1.3  &  -2.4  &  4640  &  1.2  &  -1.9  &  4515  &  0.66  &  -2.11  &  o, o, ab \\
2M21300696+1207465  &  M15  &  --  &  4709  &  0.85  &  -2.3  &  4940  &  1.5  &  -2.37  &  4656  &  1.02  &  -1.93  &  4908  &  0.83  &  -2.14  &  ab \\
2M21301049+1210061  &  M15  &  --  &  4486  &  1.02  &  -2.1  &  4470  &  1.07  &  -2.22  &  4668  &  1.41  &  -1.86  &  4562  &  1.12  &  -1.93  &  e, o, af \\
2M21304412+1211226  &  M15  &  --  &  4539  &  1.01  &  -2.08  &  4640  &  1.4  &  -2.37  &  4741  &  1.54  &  -1.88  &  4729  &  1.27  &  -1.95  &  ab \\
2M19533747+1844596  &  M71  &  High  &  4069  &  1.33  &  -0.7  &  4200  &  1.2  &  -0.83  &  4068  &  1.34  &  -0.69  &  4127  &  1.1  &  -0.8  &  p, af \\
2M19533757+1847286  &  M71  &  High  &  3906  &  1.22  &  -0.58  &  3996  &  0.88  &  -0.43  &  3918  &  1.15  &  -0.56  &  3974  &  0.84  &  -0.68  &  p, w, af \\
2M19533986+1843530  &  M71  &  High  &  4230  &  1.52  &  -0.68  &  4300  &  1.25  &  -0.7  &  4204  &  1.42  &  -0.66  &  4257  &  1.27  &  -0.74  &  p, af \\
2M19534750+1846169  &  M71  &  High  &  4259  &  1.6  &  -0.65  &  4367  &  1.55  &  -0.72  &  4270  &  1.58  &  -0.65  &  4283  &  1.25  &  -0.76  &  a, p, af, aj \\
2M19534827+1848021  &  M71  &  High  &  3979  &  1.33  &  -0.6  &  4055  &  0.8  &  -0.89  &  3968  &  1.19  &  -0.59  &  3988  &  0.79  &  -0.71  &  p, aa, af \\
2M19535064+1849075  &  M71  &  High  &  4223  &  1.71  &  -0.65  &  4291  &  1.4  &  -0.83  &  4212  &  1.55  &  -0.65  &  4226  &  1.37  &  -0.71  &  a, p, af, aj, ak \\
2M19535325+1846471  &  M71  &  High  &  3896  &  1.15  &  -0.63  &  4099  &  0.8  &  -0.89  &  3922  &  1.11  &  -0.61  &  3988  &  0.8  &  -0.74  &  a, p, aa, af \\
2M19205287+3745331  &  N6791  &  High  &  4500  &  2.58  &  0.3  &  4512  &  2.32  &  0.47  &  4520  &  2.37  &  0.32  &  4590  &  2.31  &  0.26  &  b, t \\
2M19205629+3744334  &  N6791  &  High  &  4472  &  2.53  &  0.32  &  4473  &  2.33  &  0.56  &  4456  &  2.26  &  0.34  &  4581  &  2.27  &  0.3  &  b, t \\
2M19411355+4012205  &  N6819  &  --  &  4802  &  2.51  &  0.02  &  4835  &  2.61  &  0.1  &  4805  &  2.61  &  0.01  &  4783  &  2.67  &  0.03  &  c \\
2M19413031+4009005  &  N6819  &  --  &  4079  &  1.55  &  0.04  &  4046  &  1.77  &  -0.18  &  4082  &  1.54  &  0.04  &  4139  &  1.57  &  0.04  &  v \\
2M23565751+5645272  &  N7789  &  --  &  4560  &  2.37  &  -0.01  &  4345  &  2.2  &  0.15  &  4562  &  2.36  &  -0.01  &  4626  &  2.45  &  0.05  &  ai \\
2M06293009-3116587  &  N2243  &  --  &  4595  &  2.08  &  -0.47  &  4561  &  1.9  &  -0.69  &  4603  &  2.14  &  -0.48  &  4639  &  2.08  &  -0.46  &  g, s \\
\hline

\end{longtable}

\end{scriptsize}

1. Average error from the ASPCAP DR13 parameters.

2. Average standard deviation of the mean from multiple (N$\sim$3) optical parameter sources.

3. Average statistical errors for StarNet predictions with respect to ASPCAP DR13 parameters.

Source references: (a) \citet{leep1987high}; (b) \citet{carretta2007chemical}; (c) \citet{bragaglia2001metal}; (d) \citet{carretta2009anticorrelation}; (e) \citet{minniti1993bulge}; (f) \citet{brown1991high}; (g) \citet{gratton1994}; (h) \citet{peterson1980}; (i) \citet{randich2006element}; (j) \citet{cavallo2000}; (k) \citet{lehnert1991abundances}; (l) \citet{sneden1992oxygen}; (m) \citet{sobeck2011abundances}; (n) \citet{sneden1991oxygen}; (o) \citet{sneden1997star}; (p) \citet{sneden1994oxygen}; (q) \citet{cohen1979abundances}; (r) \citet{cohen1978abundances}; (s) \citet{gratton1982chemical}; (t) \citet{gratton2006metallicity}; (u) \citet{lai2011chemical}; (v) \citet{molenda2013atmospheric}; (w) \citet{bessell1983abundance}; (x) \citet{sneden2000barium}; (y) \citet{luck1981extremely}; (z) \citet{kraft1999extremely}; (aa) \citet{cohen1980abundances}; (ab) \citet{shetrone1998keck}; (ac) \citet{kraft1992oxygen}; (ad) \citet{kraft1993oxygen}; (ae) \citet{roederer2011heavy}; (af) \citet{carretta1997abundances}; (ag) \citet{kraft1997proton}; (ah) \citet{lamb2015chemical}; (ai) \citet{pilachowski1985abundances}; (aj) \citet{Gratton1986}; (ak) \citet{pilachowski1983chemical}; (al) \citet{pilachowski1980chemical}; (am) \citet{pancino2010chemical} 
\end{landscape}

\bsp	
\label{lastpage}
\end{document}